\documentclass[pra,twocolumn,aps,showpacs]{revtex4-1}

\usepackage{epsfig}
\usepackage{graphicx}
\usepackage{amsmath}
\usepackage{amssymb}
\usepackage{enumerate}
\newcommand{\sgn}{\text{sgn}}

\usepackage{color}
\definecolor{rosso}{rgb}{1,0,0}
\definecolor{verde}{rgb}{0,1,0}
\definecolor{blue}{rgb}{0,0,1}
\definecolor{verdescuro}{rgb}{0,0.5,0.5}
\definecolor{rossoscuro}{rgb}{0.7,0.3,0}
\definecolor{bluscuro}{rgb}{0.3,0,0.7}
\definecolor{magenta}{rgb}{1,0,1}

\begin{document}

\title{Non-local equation for the superconducting gap parameter}

\author{S. Simonucci and G. Calvanese Strinati}

\affiliation{Division of Physics, School of Science and Technology \\ Universit\`{a} di Camerino, 62032 Camerino (MC), Italy \\
                 and \\ INFN, Sezione di Perugia, 06123 Perugia (PG), Italy}


\begin{abstract}
The properties are considered in detail of a non-local (integral) equation for the superconducting gap parameter, which is obtained by a coarse-graining procedure applied to the Bogoliubov-de Gennes (BdG) equations over the whole coupling-vs-temperature phase diagram associated with the superfluid phase.
It is found that the limiting size of the coarse-graining procedure, which is dictated by the range of the kernel of this integral equation, corresponds to the size of the Cooper pairs over the whole coupling-vs-temperature phase diagram up to the critical temperature, even when Cooper pairs turn into composite bosons on the BEC side of the BCS-BEC crossover.
A practical method is further implemented to solve numerically this integral equation in an efficient way, which is based on a novel algorithm for calculating the Fourier transforms.
Application of this method to the case of an isolated vortex, throughout the BCS-BEC crossover and for all temperatures in the superfluid phase, helps clarifying the nature of the length scales associated with a single vortex and the kinds of details that are in practice disposed off by the coarse-graining procedure on the BdG equations.
\end{abstract}

\pacs{74.20.Fg,03.75.Ss,05.30.Jp,74.25.Uv}
                 
\maketitle

\section{Introduction} 
\label{sec:introduction}
\vspace{-0.3cm}

Non-locality lies at the heart of the phenomenon of superconductivity and is related to the finite spatial size $\xi_{\mathrm{pair}}$ of Cooper pairs. 
The first recognition of this phenomenon came from the work by Pippard \cite{Pippard-1953,Pippard-1955}, who realised that (in a clean system) the supercurrent at a spatial point $\mathbf{r}$ is determined by a spatial average of the vector potential $\mathbf{A}$ over a neighboring region with the size of a Cooper pair.

When solving for the gap equation in inhomogeneous situations to obtain the gap parameter $\Delta(\mathbf{r})$, a second length scale (the healing length $\xi$) enters the problem  to describe the typical distance over which $\Delta(\mathbf{r})$ tends to recover its bulk value in the presence of a localized perturbation.
Both lengths $\xi_{\mathrm{pair}}$ and $\xi$ depend on the coupling value of the inter-particle interaction \cite{Pistolesi-1994,Pistolesi-1996}, which gives rise to Cooper pairs and their condensation to begin with, as well as on temperature \cite{Palestini-2014}.
The relative behaviour of $\xi$ with respect to $\xi_{\mathrm{pair}}$ is expected to determine the relevance of the local vs non-local behaviour of $\Delta(\mathbf{r})$ with respect to its surrounding values.

Gor'kov \cite{Gorkov-1959} first realized the importance of distinguishing between the two lengths $\xi_{\mathrm{pair}}$ and $\xi$, in the process of deriving the Ginzburg-Landau (GL) equation from the coupled differential equations for the normal and anomalous single-particle temperature Green's functions (or, alternatively, from the integral version of the 
Bogoliubov-de Gennes (BdG) equations \cite{deGennes-1966}).
Although Gor'kov's derivation applies to the weak-coupling limit whereby $\xi_{\mathrm{pair}}$ is much larger than the inter-particle distance (as specified by the inverse of the Fermi wave vector $k_{F} = (3 \pi^{2} n)^{1/3}$ where $n$ is the particle density), the local (differential) GL equation could be retrieved close to the critical temperature $T_{c}$ where $\xi \gg \xi_{\mathrm{pair}}$.
In the opposite case of strong coupling, when Cooper pairs turn into composite bosons with size $\xi_{\mathrm{pair}} \ll k_{F}$, the condition $\xi \gg \xi_{\mathrm{pair}}$ can be realized also at zero temperature.
In this case, it is the local (differential) Gross-Pitaevskii (GP) equation for composite bosons that can be derived from the BdG equations, as shown in Ref.~\cite{Pieri-2003}.

The way $\xi$ and $\xi_{\mathrm{pair}}$ evolve with respect to each other at zero temperature characterises  the BCS-BEC crossover, whereby $\xi=\xi_{\mathrm{pair}}$ in the BCS limit and $\xi \gg \xi_{\mathrm{pair}}$ in the BEC limit \cite{Pistolesi-1994,Pistolesi-1996}. 
In practice, the crossover between these two limits can be obtained by varying the inter-particle attraction, in such a way that the system evolves from a BCS state where pairs of (opposite spin) fermions are described by Fermi statistics, to a BEC state where two-fermion dimers (or composite bosons) are described by Bose statistics. 
A substantial amount of theoretical work \cite{Eagles-1969,Leggett-1980,NSR-1985,Randeria-1989-90} had preceded the explicit experimental realization of the BCS-BEC crossover with ultra-cold Fermi gases \cite{Jin-2003,Hulet-2003,Grimm-2003,Jin-2004}. 
In these systems, the attraction between opposite-spin fermions can be taken to be of zero range in space and instantaneous in time,
conditions that will be assumed to hold in the rest of this paper in line with the original Galitskii approach for a dilute Fermi system \cite{Galitskii-1958}. 
In condensed matter, where the inter-particle interaction can have more complicated forms, on the other hand, consideration of the BCS-BEC crossover was originally suggested by the fact that in high-$T_{c}$ superconductors the product $k_{F}\xi_{\mathrm{pair}}$ is of order unity, corresponding to the ``unitary'' regime which is intermediate between the BCS and BEC limits
\cite{Pistolesi-1994}.  
Growing evidence for the occurrence of this crossover has lately emerged also in two-band superconductors with iron-based materials \cite{Chubukov-2016}. 

In the context of the BdG and related equations, recently a method was devised to obtain a non-linear differential equation for the gap parameter $\Delta(\mathbf{r})$ by performing a suitable spatial coarse graining of the BdG equations, which deals with the \emph{smoothness} of the spatial variations of the magnitude and phase of $\Delta(\mathbf{r})$ on a different footing for smoothing out short-range details of the gap parameter \cite{Simonucci-2014}.
This equation (referred to as a local phase density approximation (LPDA) to the BdG equations) was found to recover \emph{both} the GL equation in weak coupling close $T_{c}$ and the GP equation in strong coupling at low temperature.
In Ref.~\cite{Simonucci-2014} the LPDA equation was applied at any temperature in the superfluid phase throughout the BCS-BEC crossover, for the test case of an isolated vortex for which also an accurate numerical solution of the BdG equations is available to compare with \cite{Simonucci-2013}.
This test led to an extremely good agreement between the LPDA and BdG calculations essentially for all couplings and temperatures, with the exception of the BCS (weak-coupling) regime at low temperature where deviations between the two calculations have emerged.
In Ref.~\cite{Simonucci-2014} the reason for this discrepancy was attributed to the fact that, in this regime, the vortex size (or healing length $\xi$) becomes comparable with the Cooper pair size 
$\xi_{\mathrm{pair}}$ \cite{Pistolesi-1994,Pistolesi-1996}, thereby questioning the validity of a local (differential) approach like the LPDA equation in this restricted regime of coupling and temperature.
Later, the LPDA approach was successfully applied to generate in a self-consistent way large arrays of vortices throughout the BCS-BEC crossover \cite{Simonucci-2015}, which can be produced by setting an ultra-cold trapped Fermi gas into rotation.
In this way, it was possible to account for the experimental data that had provided the first direct evidence of the superfluid phase in these systems \cite{Zwierlein-2005}.

Although the end result of the coarse-grained derivation of Ref.~\cite{Simonucci-2014} was the \emph{local} (differential) LPDA equation, a \emph{non-local} (integral) version of the LPDA equation 
(that can accordingly be referred to as the NLPDA equation) was also reported in that reference, at an intermediate step between the original BdG equations and the LPDA equation.
The (integral) NLPDA equation for the gap parameter contains a non-local kernel that depends on the superfluid gap itself in a highly non-linear way.
However, the (non-local) NLPDA equation was not further examined in Ref.~\cite{Simonucci-2014}, where the attention was concentrated only on the (local) LPDA equation.
Examination of the (non-local) NLPDA equation can also be of interest in itself, since it can give access to problems that are difficult to deal with using the (local) LPDA equation.

Purpose of this paper is to consider in detail the non-local (integral) NLPDA equation for the gap parameter and study the properties of its kernel, in order to determine its spatial range for all couplings 
and temperatures in the superfluid phase throughout the BCS-BEC crossover.
This turns out be a non-trivial task, especially in the weak-coupling (BCS) regime at low temperature where the kernel shows rapid and slowly decaying spatial oscillations. 
This study eventually enables us to identify the spatial extent of the ``granularity'' associated with the coarse-graining procedure on which the NLPDA (and, as a consequence, the LPDA) equation rests,
as well as to understand the reason for the failure of the LPDA equation in weak coupling at low temperature mentioned above.
We will find that, for all couplings throughout the BCS-BEC crossover and temperatures in the superfluid phase, the spatial range of this kernel coincides with the Cooper pair size $\xi_{\mathrm{pair}}$, 
a quantity which was independently determined in Ref.~\cite{Palestini-2014} by analyzing the pair correlation function for opposite-spin fermions.
In this context, a method will also be implemented for solving numerically the (integral) NLPDA equation in an efficient way.
This method will be explicitly utilized to study an isolated vortex and to compare the results with the solutions of both the LPDA \cite{Simonucci-2014} and BdG \cite{Simonucci-2013} calculations, for all couplings throughout the BCS-BEC crossover and temperatures in the superfluid phase.
This test calculation will highlight the length scales associated with an isolated vortex by the three (BdG, LPDA, and NLPDA) calculations, a result that will be especially instructive in weak coupling at low temperature.

The paper is organized as follows.
In Section~\ref{sec:Kernel-NLPDA} the properties of the kernel of the NLPDA equation are studied in detail, both in wave-vector and real space, and the spatial extent of this kernel is determined for all couplings throughout the BCS-BEC crossover and temperatures in the superfluid phase.
Knowledge of these properties will also enable us to set up an efficient strategy for the numerical solution of the NLPDA equation.
This will be explicitly done in Section~\ref{sec:solving-NLPDA-equation} for the case of an isolated vortex embedded in an infinite superfluid, for all couplings throughout the BCS-BEC crossover and temperatures in the superfluid phase.
The range validity of the (local) LPDA equation will also be explicitly determined from an analysis of the (non-local) NLPDA equation from which it is derived.
Section~\ref{sec:conclusions} gives our conclusions.
In Appendix~\ref{sec:appendix-A} a model kernel is introduced in wave-vector space, whose analytic solution in real space will help us to identify the origin of the spatial oscillations of the kernel of the NLPDA equation that show up in weak coupling at low temperature.
In Appendix~\ref{sec:appendix-B} a method is implemented for solving numerically the NLPDA equation, by devising a novel approach for calculating the Fourier transforms from real to wave-vector space and vice versa.

\vspace{-0.2cm}
\section{The kernel of the Non-Local NLPDA gap equation} 
\label{sec:Kernel-NLPDA}
\vspace{-0.3cm}

The following non-local (integral) gap equation
\begin{small}
\begin{equation}
- \frac{m}{4 \pi a_{F}} \, \Delta(\mathbf{r}) =  \! \int \! d \mathbf{R} \,\, \Delta(\mathbf{R})  
\int \!  \frac{d\mathbf{Q}}{\pi^{3}} \, e^{2 i \mathbf{Q} \cdot (\mathbf{r} - \mathbf{R})} \, K^{\mathbf{A}}(\mathbf{Q}|\mathbf{r}) 
\label{non-local-LPDA-equation}
\end{equation}
\end{small}

\noindent
was obtained in Ref.~\cite{Simonucci-2014} at an intermediate step in the process of deriving the local (differential) LPDA equation by a coarse-graining procedure of the BdG equations.
The kernel of this equation reads:
\begin{small}
\begin{equation}
K^{\mathbf{A}}(\mathbf{Q}|\mathbf{r}) \! = \! \! \int \! \frac{d\mathbf{k}}{(2 \pi)^{3}} 
\left\{ \frac{ 1 - 2 \, f_{F}(E_{+}^{\mathbf{A}}(\mathbf{k};\mathbf{Q}|\mathbf{r}))}
{2 E^{\mathbf{A}}(\mathbf{k};\mathbf{Q}|\mathbf{r})} - \frac{m}{\mathbf{k}^{2}} \right\} 
\label{Kernel-Q} 
\end{equation}
\end{small}

\noindent
where $m$ is the fermion mass, $a_{F}$ the scattering length of the two-fermion problem,
$f_{F}(E) = \left( e^{E/(k_{B}T)} + 1 \right)^{-1}$ the Fermi function at temperature $T$ ($k_{B}$ being the Boltzmann constant),
\begin{footnotesize}
\begin{eqnarray}
E_{\pm}^{\mathbf{A}}(\mathbf{k};\mathbf{Q}|\mathbf{r}) & = &
\sqrt{\left( \frac{\mathbf{k}^{2}}{2m} + \frac{\mathbf{Q}^{2}}{2m} - \bar{\mu}(\mathbf{r}) - \frac{\mathbf{A}(\mathbf{r})}{m} \cdot \mathbf{Q} \right)^{2} + |\Delta(\mathbf{r})|^{2}} 
\nonumber \\
& \pm & \frac{\mathbf{k}}{m} \cdot (\mathbf{Q}-\mathbf{A}(\mathbf{r})) \, ,
\label{definition-E-pm}
\end{eqnarray}
\end{footnotesize}

\noindent
and $2 E^{\mathbf{A}}(\mathbf{k};\mathbf{Q}|\mathbf{r}) = E_{+}^{\mathbf{A}}(\mathbf{k};\mathbf{Q}|\mathbf{r}) + E_{-}^{\mathbf{A}}(\mathbf{k};\mathbf{Q}|\mathbf{r})$.
In the above expressions, $\mathbf{A}(\mathbf{r})$ is the vector potential, $\bar{\mu}(\mathbf{r}) = \mu - V(\mathbf{r}) - \mathbf{A}(\mathbf{r})^{2}/(2m)$ the local chemical potential in the presence 
of an external (trapping) potential $V(\mathbf{r})$, and $|\Delta(\mathbf{r})|$ the magnitude of the local gap parameter.

Apart from its role as an intermediate step in deriving the local (differential) LPDA equation, no further consideration was given in Ref.~\cite{Simonucci-2014} to the non-local (integral) NLPDA equation 
(\ref{non-local-LPDA-equation}), since all attention was concentrated on the LPDA equation.
Here, our interest is to focus directly on the NLPDA equation (\ref{non-local-LPDA-equation}) and study in detail the properties of its kernel $K^{\mathbf{A}}(\mathbf{Q}|\mathbf{r}) $ as well as of its Fourier transform
\begin{equation}
K^{\mathbf{A}}(\mathbf{R}|\mathbf{r})  = \int \! \frac{d\mathbf{Q}}{\pi^{3}} \, e^{2 i \mathbf{Q} \cdot \mathbf{R}} \, K^{\mathbf{A}}(\mathbf{Q}|\mathbf{r}) \, ,
\label{Kernel-R}
\end{equation}
\noindent
aiming at determining its spatial range in real space.
To this end, it will be sufficient to consider the pivotal case with $\mathbf{A}(\mathbf{r})=0$, $V(\mathbf{r})=0$, and $|\Delta(\mathbf{r})| \rightarrow \Delta$, where $\Delta$ is the uniform mean-field value 
of the gap parameter for the homogeneous system \cite{Physics-Reports-2017}.
Accordingly, we set $K^{\mathbf{A}}(\mathbf{R}|\mathbf{r}) \rightarrow K^{\mathbf{A=0}}(\mathbf{R}) \equiv K(\mathbf{R})$ to simplify the notation.

This analysis will be carried out for given temperature in the superfluid phase and coupling throughout the BCS-BEC crossover, whereby the coupling parameter $(k_F a_F)^{-1}$ ranges from 
$(k_F\, a_F)^{-1} \lesssim -1$ in the weak-coupling (BCS) regime when $a_F < 0$, to $(k_F\, a_F)^{-1} \gtrsim +1$ in the strong-coupling (BEC) regime when 
$a_F > 0$, across the unitary limit (UL) when $|a_F|$ diverges.    
In practice, the ``crossover region'' of most interest is approximately limited by the interval $-1 \lesssim (k_F\, a_F)^{-1} \lesssim +1$.

\vspace{0.05cm}
\begin{center}
{\bf A. Properties of the kernel $K(\mathbf{Q})$}
\end{center}
\vspace{-0.2cm}

We begin by considering a number of properties of the kernel $K(\mathbf{Q})$ in $\mathbf{Q}$-space, which is given by the expression (\ref{Kernel-Q}) with the further account of the above provisions.

\vspace{0.1cm}
\noindent
{\bf Spherical symmetry.}
When $\mathbf{A}(\mathbf{r})=0$, the kernel $K(\mathbf{Q})$ depends only on $Q = |\mathbf{Q}|$.
This is because the transformation $\mathbf{Q} \rightarrow \mathcal{R} \mathbf{Q}$ where $\mathcal{R}$ is a three-dimensional rotation can be compensated by an analogous rotation
$\mathbf{k} \rightarrow \mathcal{R} \mathbf{k}$ of the integration variable in Eq.~(\ref{Kernel-Q}).
We thus write $K(Q)$ in place of $K(\mathbf{Q})$ (and, correspondingly, $K(R)$ in place of $K(\mathbf{R})$ where $R = |\mathbf{R}|$).

\vspace{0.1cm}
\noindent
{\bf Non analyticity at $\mathbf{T=0}$.}
At zero temperature, the Fermi function in Eq.~(\ref{Kernel-Q}) is non-vanishing as soon as its argument becomes negative. 
When this occurs, the Fermi function has a step singularity, which reflects itself in a ``kink'' in the kernel $K(Q)$ as a function of $Q$ at a critical value $Q_{c}$.
This kink, in turn, considerably affects the large-$R$ behaviour of the Fourier transform $K(R)$, to be discussed below.
The value of $Q_{c}$ is determined as follows.
When $\mathbf{A}(\mathbf{r})=0$, $V(\mathbf{r})=0$, and $|\Delta(\mathbf{r})| \rightarrow \Delta$, the argument of the Fermi function in Eq.~(\ref{Kernel-Q})
\begin{small}
\begin{equation}
E_{+}(\mathbf{k};\mathbf{Q}) =
\sqrt{\left( \frac{\mathbf{k}^{2}}{2m} + \frac{\mathbf{Q}^{2}}{2m} - \mu \right)^{2} + |\Delta|^{2}} \, + \, \frac{\mathbf{k}}{m} \cdot \mathbf{Q} 
\label{definition-E-plus}
\end{equation}
\end{small}

\noindent
first approaches zero for $\mathbf{k} \| \mathbf{Q}$, such that $\mathbf{k} \cdot \mathbf{Q} = - k \, Q$ with $k = |\mathbf{k}|$.
Setting $x = Q^{2}/(2m)$, the condition for the expression (\ref{definition-E-plus}) to vanish becomes
\begin{equation}
x^{2} - 2 \, x \, (\varepsilon_{k} + \mu) + (\varepsilon_{k} - \mu)^{2} + |\Delta|^{2} = 0
\label{quadratic-equation}
\end{equation}
\noindent
where $\varepsilon_{k} = k^{2}/(2m)$, whose solutions
\begin{equation}
x_{\pm}(k) = \varepsilon_{k} + \mu \pm \sqrt{4 \, \varepsilon_{k} \, \mu - |\Delta|^{2}}
\label{solutions}
\end{equation}
\noindent
are both acceptable provided $4 \varepsilon_{k} \mu \ge |\Delta|^{2}$ for $\mu > 0$.
However, only $x_{-}(k)$ attains a minimum for $\varepsilon_{k} = \mu + |\Delta|^{2}/(4 \mu)$, in correspondence to which 
$x_{-}(k) |_{\mathrm{min}} \equiv Q_{c}^{2}/(2m) = |\Delta|^{2}/(4 \mu)$.
When $\mu<0$, on the other hand, the argument of the Fermi function in Eq.~(\ref{Kernel-Q}) never vanishes and the kernel $K(Q)$ is a smooth function of $Q$.
[Recall in this context that, at the mean-field level, the zero-temperature chemical potential changes its sign at about the coupling value $(k_{F} a_{F})^{-1} = 0.55$.]
\begin{figure}[t]
\begin{center}
\includegraphics[width=8.5cm,angle=0]{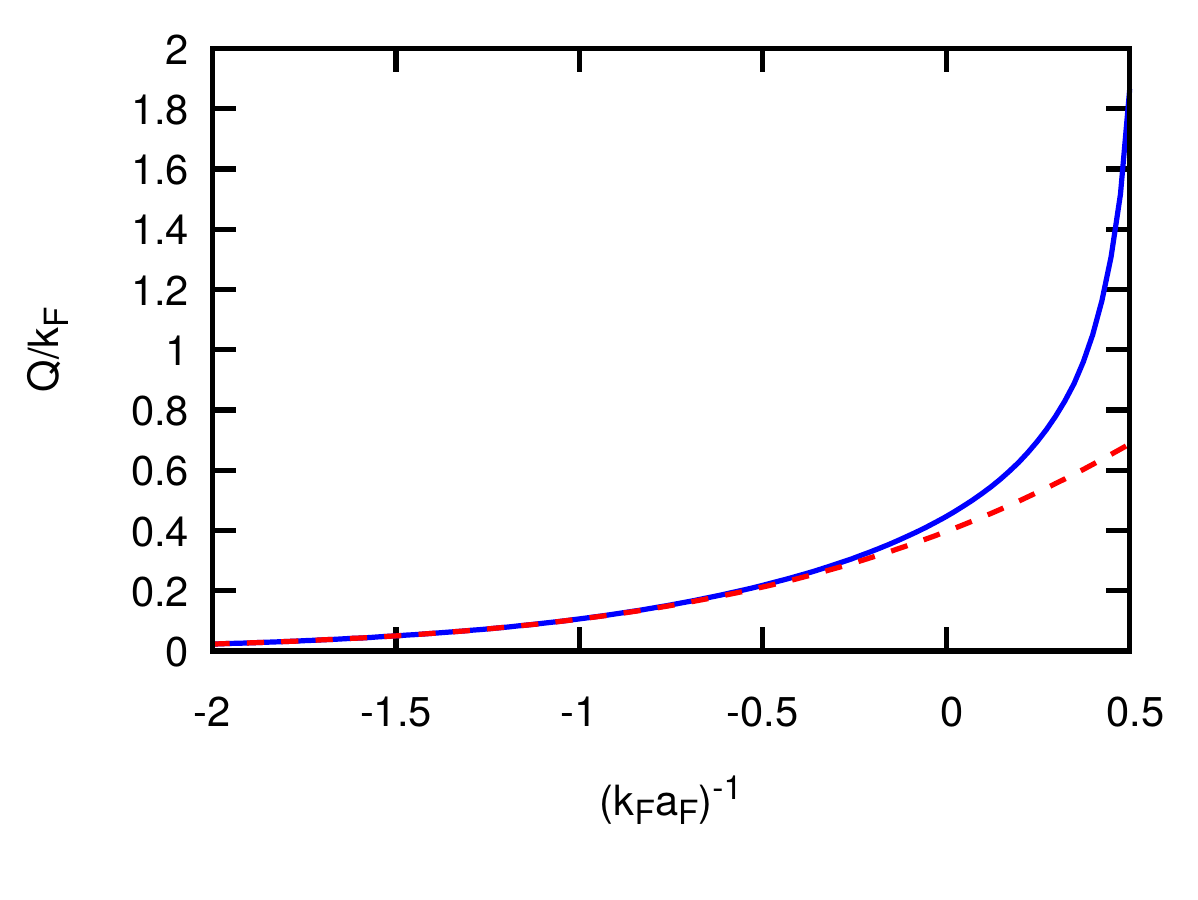}
\caption{(Color online) Coupling dependence of the critical wave vector $Q_{c}$ at which the kink of the kernel $K(Q)$ occurs at $T=0$ (full line) and of the critical Landau wave vector 
                                    $Q_{c}^{\mathrm{L}}$ given by the expression (\ref{Q_c-Landau-criterion}) (dashed line). Both expressions are obtained with the mean-field values of $\Delta$ and $\mu$ at $T=0$.}
\label{Figure-1}
\end{center}
\end{figure}
The critical value $Q_{c}$ for the kernel $K(Q)$ differs, in general, from the critical value $Q_{c}^{\mathrm{L}}$ of the Landau criterion for superfluidity associated with pair-breaking excitations, which is given by the expression (cf., e.g. Section 4.6 of Ref.~\cite{Spuntarelli-2010}):
\begin{equation}
\frac{(Q_{c}^{\mathrm{L}})^{2}}{m} = \sqrt{ \mu^{2} +|\Delta|^{2} } - \mu \, .
\label{Q_c-Landau-criterion}
\end{equation}
\noindent
The value of $Q_{c}^{\mathrm{L}}$ approaches $Q_{c}$ only asymptotically in the weak-coupling (BCS) limit when $\mu > 0$ and $|\Delta| \ll \mu$.

Figure~\ref{Figure-1} shows the dependence on the coupling $(k_{F} a_{F})^{-1}$ of the two critical wave vectors $Q_{c}$ and $Q_{c}^{\mathrm{L}}$, which are obtained using the mean-field values of $\Delta$ and $\mu$ at zero temperature.

At finite temperature, on the other hand, the Fermi function in Eq.~(\ref{Kernel-Q}) is a smooth function of its argument, resulting in a smooth dependence of the kernel $K(Q)$ on $Q$.
This feature, in turn, will make the dependence of $K(R)$ on $R$ less problematic than at zero temperature.

\vspace{0.2cm}
\noindent
{\bf Small-$\mathbf{Q}$ behaviour.}
In the homogeneous case with a uniform gap parameter, the gap equation (\ref{non-local-LPDA-equation}) reduces to the form:
\begin{equation}
- \frac{m}{4 \pi a_{F}} =  K(Q=0) \equiv \mathcal{I}_{0}
\label{non-local-LPDA-equation-uniform}
\end{equation}
\noindent
with the notation $\mathcal{I}_{0}$ introduced in Ref.~\cite{Simonucci-2014} (in which we now set $\mathbf{A}(\mathbf{r})=0$, $V(\mathbf{r})=0$, and $|\Delta(\mathbf{r})| \rightarrow \Delta$).
This condition identifies the value of $K(Q=0)$ at self-consistency.
In particular, one sees that $K(Q=0)$ is positive (negative) on the BCS (BEC) side of the crossover where $a_{F} < 0$ ($a_{F} > 0$), and vanishes at unitarity where $a_{F} = \pm \infty$.
Near $Q=0$, $K(Q)$ decreases quadratically in $Q$ with coefficient
\begin{equation}
\left. \frac{ d^{2} K(Q)}{d Q^{2}} \right|_{Q=0} = - \frac{2 \, \mathcal{I}_{1}}{m}
\label{second-derivative-Q}
\end{equation}
where the notation $\mathcal{I}_{1}$ was also introduced in Ref.~\cite{Simonucci-2014} (in which we again set $\mathbf{A}(\mathbf{r})=0$, $V(\mathbf{r})=0$, and 
$|\Delta(\mathbf{r})| \rightarrow \Delta$).
This expansion up to quadratic order in $Q$ about $Q=0$ was utilized in Ref.~\cite{Simonucci-2014} to derive the (differential) LPDA equation from the (integral) NLPDA equation.

It was also shown in Ref.~\cite{Simonucci-2014} that, at zero temperature, both $\mathcal{I}_{0}$ and $\mathcal{I}_{1}$ can be calculated analytically in terms of elliptic integrals according 
to the method of Ref.~\cite{Marini-1998}.
A plot of $\mathcal{I}_{1}$ as a function of the coupling $(k_{F} a_{F})^{-1}$ (given in Fig.~7 of Ref.~\cite{Simonucci-2014}) shows that $\mathcal{I}_{1}$ is a monotonically decreasing function of coupling 
from the BCS to the BEC limits.
These results were also used in Ref.~\cite{Brand-2014} to apply the LPDA equation to the study of the snake instability of dark solitons.

\vspace{0.2cm}
\noindent
{\bf Large-$\mathbf{Q}$ behaviour.}
For large values of $|\mathbf{Q}|$, the term $|\Delta|^{2}$ in Eq~.(\ref{definition-E-plus}) can be neglected irrespective of the sign of $\mu$, such that  
$E(\mathbf{k};\mathbf{Q}) \simeq (\mathbf{k}^{2} + \mathbf{Q}^{2})/(2m) - \mu$.
In addition, $E_{+}(\mathbf{k};\mathbf{Q}) \simeq (\mathbf{k} + \mathbf{Q})^{2}/(2m) - \mu$ becomes negative only for positive values of $\mu$ when $|\mathbf{k} + \mathbf{Q}| \le \sqrt{2 m \mu}$.
At zero temperature, the contribution to the kernel $K(\mathbf{Q})$ for large $|\mathbf{Q}|$ originating from the presence of the Fermi function then reads:
\begin{small}
\begin{eqnarray}
& & \int \!\! \frac{d\mathbf{k}}{(2 \pi)^{3}} \frac{ f_{F}(E_{+}(\mathbf{k};\mathbf{Q})) }{ E(\mathbf{k};\mathbf{Q}) } 
\simeq  \frac{m}{\mathbf{Q}^{2}} \! \int \!\! \frac{d\mathbf{k}}{(2 \pi)^{3}} \, \Theta \left( \sqrt{2 m \mu} - |\mathbf{k} + \mathbf{Q}| \right) 
\nonumber \\
& = & \frac{m \, (2 m \mu)^{3/2}}{6 \, \pi^{2} \, \mathbf{Q}^{2}} 
\label{large_Q-Fermi_function}
\end{eqnarray}
\end{small}

\noindent
where $\Theta(x)$ is the Heaviside unit step function of argument $x$.
This contribution is sub-leading with respect to the remaining part of the integral that defines the kernel $K(\mathbf{Q})$ for large $|\mathbf{Q}|$, namely,
\begin{small}
\begin{eqnarray}
& & \int \!\! \frac{d\mathbf{k}}{(2 \pi)^{3}} \! \left\{ \frac{1}{2 E(\mathbf{k};\mathbf{Q})} - \frac{m}{\mathbf{k}^{2}} \right\} \simeq
\int \!\! \frac{d\mathbf{k}}{(2 \pi)^{3}} \! \left\{ \frac{m}{ \mathbf{k}^{2} + \mathbf{Q}^{2} } - \frac{m}{\mathbf{k}^{2}} \right\} 
\nonumber \\
& = & - \frac{m \, \mathbf{Q}^{2}}{2 \, \pi^{2}} \!\! \int_{0}^{\infty} \!\! \frac{dk}{k^{2}+\mathbf{Q}^{2}} = - \frac{m \, |\mathbf{Q}|}{4 \, \pi} \, .
\label{large_Q-remaining_part}
\end{eqnarray}
\end{small}

\noindent
The expression (\ref{large_Q-remaining_part}) thus gives the leading contribution to $K(\mathbf{Q})$ for large $|\mathbf{Q}|$, \emph{irrespective of coupling}.
This asymptotic result remains valid even at finite temperature, provided that $\mathbf{Q}^{2}/m \gg k_{B} T$.

\begin{figure}[h]
\begin{center}
\includegraphics[width=7.5cm,angle=0]{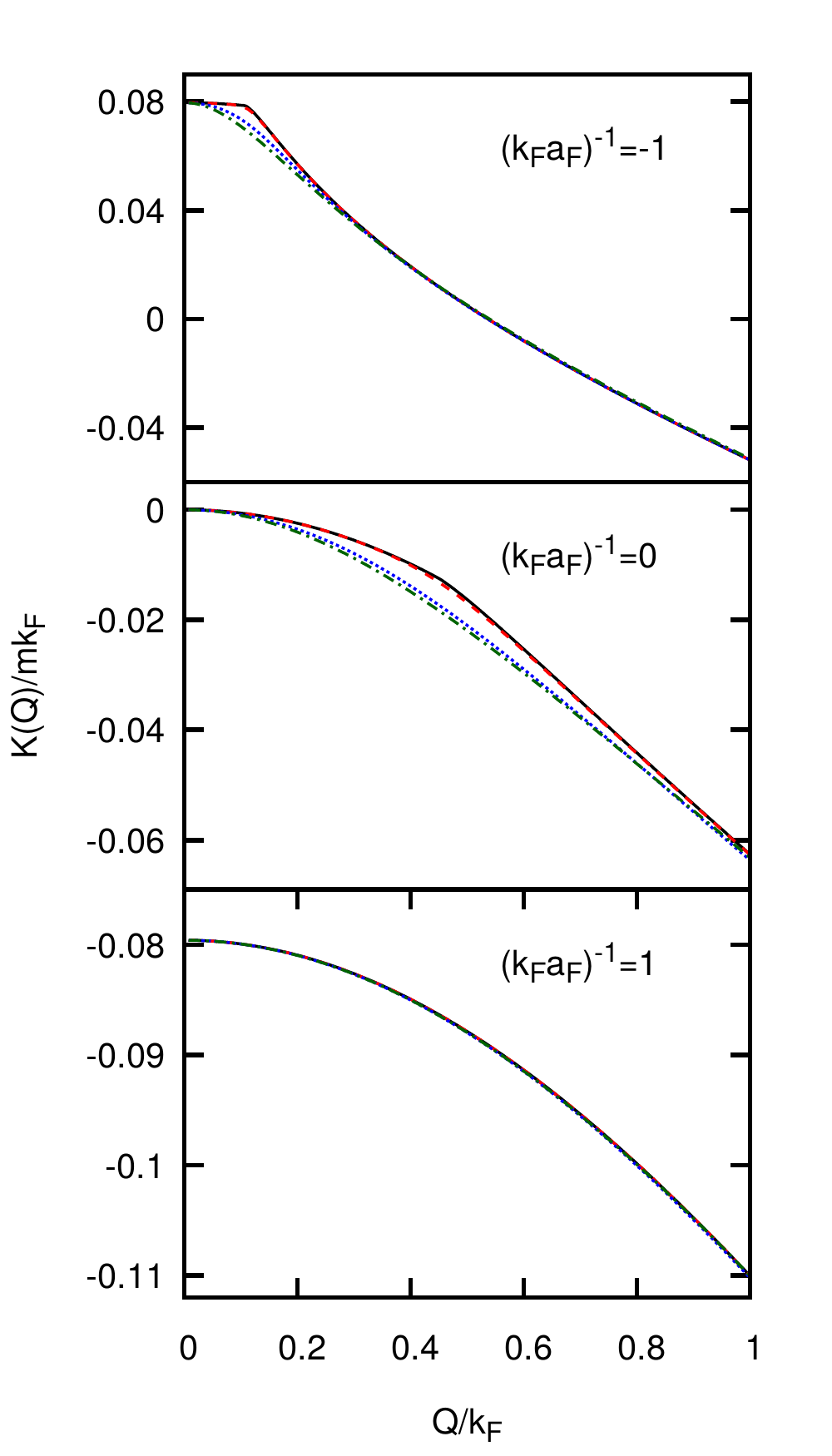}
\caption{(Color online) Wave-vector dependence of the kernel $K(Q)$ for various couplings and temperatures: $T=0$ (full line); $T=0.1T_{c}$ (dashed line); 
                                    $T=0.5T_{c}$ (dotted line); $T=0.95T_{c}$ (dashed-dotted line). In each case, the critical temperature $T_{c}$ refers to the giving coupling.}
\label{Figure-2}
\end{center}
\end{figure}

For later convenience, we identify an \emph{asymptotic kernel} defined for \emph{all} values of $\mathbf{Q}$ by the expression (\ref{large_Q-remaining_part}), namely,
\begin{equation}
K_{\infty}(\mathbf{Q}) =  - \frac{m \, |\mathbf{Q}|}{4 \, \pi}
\label{asymptotic-kernel}
\end{equation}
\noindent
irrespective of coupling and temperature.

\vspace{0.2cm}
\noindent
{\bf Angular integration.}
At any temperature, the $\mathbf{k}$-integration that enters the definition of the kernel $K(Q)$ can be reduced to a one-dimensional numerical integration over $k = |\mathbf{k}|$,
by performing analytically the integration over the angle $\hat{\mathbf{k}}$.
This is done with the use of the standard integral (where $b \ne 0$ and $\lambda \ne 0$):
\begin{equation}
\int dx \,\, \frac{1}{a \, e^{\lambda \, x} \, + \, b} = \frac{1}{b} \left[ x - \frac{1}{\lambda} \ln \left( a \, e^{\lambda \, x}  \, + \, b \right)  \right] \, .
\label{standard-integral}
\end{equation}
\noindent
From Eq.~(\ref{Kernel-Q}) one then obtains the expression:
\begin{footnotesize}
\begin{eqnarray}
& & K(Q) = \int_{0}^{\infty} \! \frac{dk \, k^{2}}{(2 \pi)^{2}} 
\label{K-one-dimensional-integral} \\ 
& &  \times \left\{ \frac{1}{E(k;Q)} \left[ \frac{m k_{B} T}{k \, Q} \ln \left( \frac{e^{(E(k;Q) + kQ/m)/k_{B}T} + 1}{e^{(E(k;Q) - kQ/m)/k_{B}T} + 1} \right) - 1 \right] - \frac{2m}{k^{2}} \right\}
\nonumber
\end{eqnarray}
\end{footnotesize}

\noindent
where $E(k;Q) = \sqrt{\left( \frac{k^{2} + Q^{2}}{2m} - \mu \right)^{2} + |\Delta|^{2}} $ as before.

\begin{figure}[t]
\begin{center}
\includegraphics[width=8.5cm,angle=0]{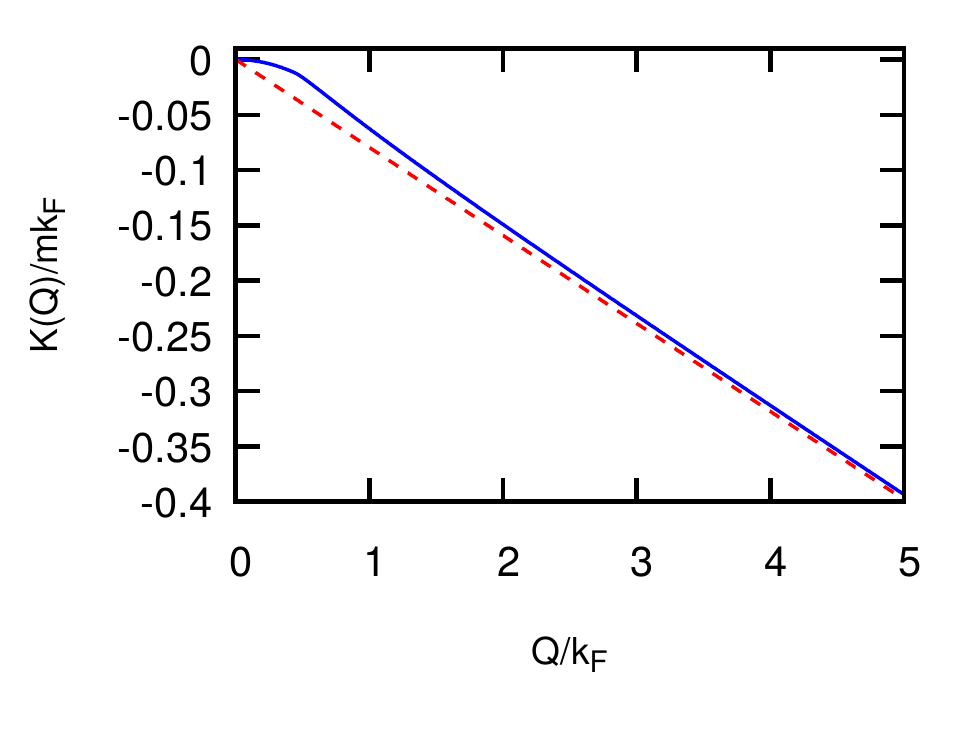}
\caption{(Color online) Wave-vector dependence of the kernel $K(Q)$ over an extended range of $Q$ at unitarity and zero temperature (full line), showing the convergence to the asymptotic kernel            
                                     (\ref{asymptotic-kernel}) (dashed line).}
\label{Figure-3}
\end{center}
\end{figure}

\vspace{0.2cm}
\noindent
{\bf Profile of $\mathbf{K(Q)}$ for various couplings and temperatures.}
Figure~\ref{Figure-2} shows the wave-vector dependence of the kernel $K(Q)$ calculated numerically from the expression (\ref{K-one-dimensional-integral}) in the range $Q \le k_{F}$, 
for various couplings and temperatures from $T=0$ up to close to $T_{c}$. 
These plots confirm the downward quadratic dependence of $K(Q)$ near $Q=0$ for all couplings and temperatures, as well as the presence of a kink at $Q = Q_{c}$ at zero temperature for coupling values before the chemical potential changes its sign (i.e., for $(k_{F} a_{F})^{-1} \le 0.55$).
One notes further from this figure that, as soon that the Fermi function in Eq.~(\ref{Kernel-Q}) becomes smooth for increasing temperature, the kink singularity in $K(Q)$ is also smoothed out.

In addition, Fig.~\ref{Figure-3} shows the wave-vector dependence of the kernel $K(Q)$ over a more extended range of Q in the case of unitarity at zero temperature, evidencing how $K(Q)$ converges, in practice, for large enough $Q$ to its asymptotic expression (\ref{large_Q-remaining_part}) (or, else, to the asymptotic kernel (\ref{asymptotic-kernel})).

\begin{figure}[t]
\begin{center}
\includegraphics[width=7.2cm,angle=0]{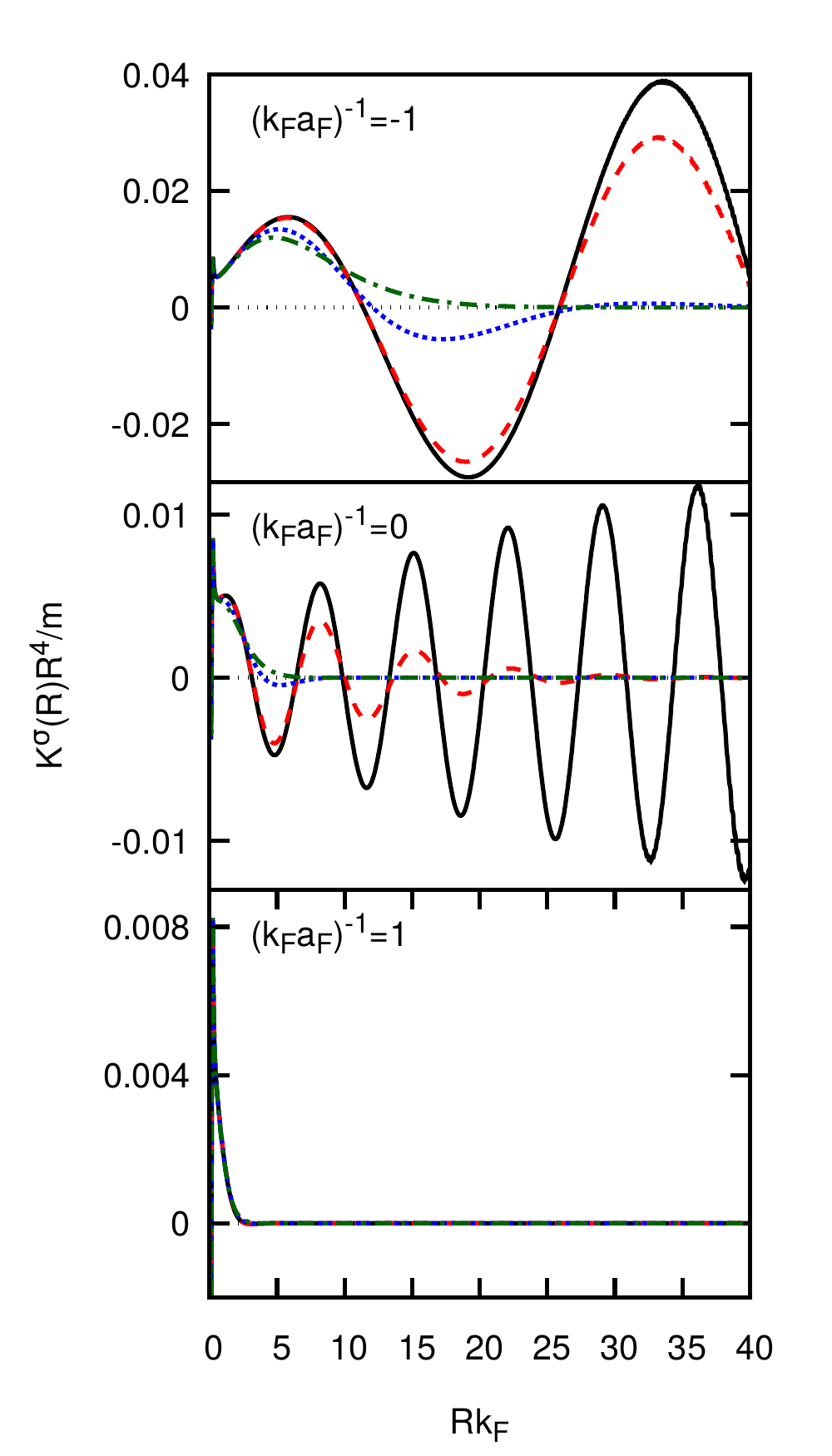}
\caption{(Color online) Radial profile of the kernel $K^{\sigma}(R)$ for various couplings and temperatures: $T=0$ (full line); $T=0.1T_{c}$ (dashed line); 
                                    $T=0.5T_{c}$ (dotted line); $T=0.95T_{c}$ (dashed-dotted line). 
                                    In each case, the critical temperature $T_{c}$ refers to the given coupling, while the value $\sigma=10 k_{F}$  is common to all curves. 
                                    To make the difference among the various curves more visible, $K^{\sigma}(R)$ is multiplied by $R^{4}$.}
\label{Figure-4}
\end{center}
\end{figure}

\vspace{0.05cm}
\begin{center}
{\bf B. Properties of the modified kernel $K^{\sigma}(\mathbf{R})$}
\end{center}
\vspace{-0.2cm}

The asymptotic behaviour (\ref{large_Q-remaining_part}) of the kernel $K(\mathbf{Q})$ suggests us that, for the later purpose of calculating the Fourier transform $K(\mathbf{R})$, it is convenient 
to \emph{interpret the kernel $K(\mathbf{Q})$ in the sense of distributions} \cite{Lighthill-1958}.
This is done by introducing a test function of the Gaussian form $e^{-\mathbf{Q}^{2} / \sigma^{2}}$, that can be included in the kernel itself by the definition:
\begin{equation}
K^{\sigma}(\mathbf{Q}) = K(\mathbf{Q}) \, e^{-\mathbf{Q}^{2} / \sigma^{2}} \, .
\label{kernel-new-definition}
\end{equation}
\noindent
It is further understood that the limit $\sigma \rightarrow \infty$ will be taken (at least formally) only \emph{after} the Fourier transform 
\begin{equation}
K^{\sigma}(\mathbf{R}) = \int \! \frac{d\mathbf{Q}}{\pi^{3}} \, e^{2 i \mathbf{Q} \cdot \mathbf{R}} \, K^{\sigma}(\mathbf{Q})
\label{Kernel-R-simplified}
\end{equation}
\noindent
will be calculated.
Although we shall consider values of $\sigma$ up to $40 k_{F}$, it will turn out that $\sigma = 20 k_{F}$ (or even less) will be sufficient for most purposes.
Similarly, a related definition
\begin{equation}
K_{\infty}^{\sigma}(\mathbf{Q})  = K_{\infty}(\mathbf{Q}) e^{-\mathbf{Q}^{2} / \sigma^{2}}
\label{asymptotic-kernel-new-definition}
\end{equation}
\noindent
is also introduced for the asymptotic kernel (\ref{asymptotic-kernel}). 

\begin{figure}[t]
\begin{center}
\includegraphics[width=6.5cm,angle=0]{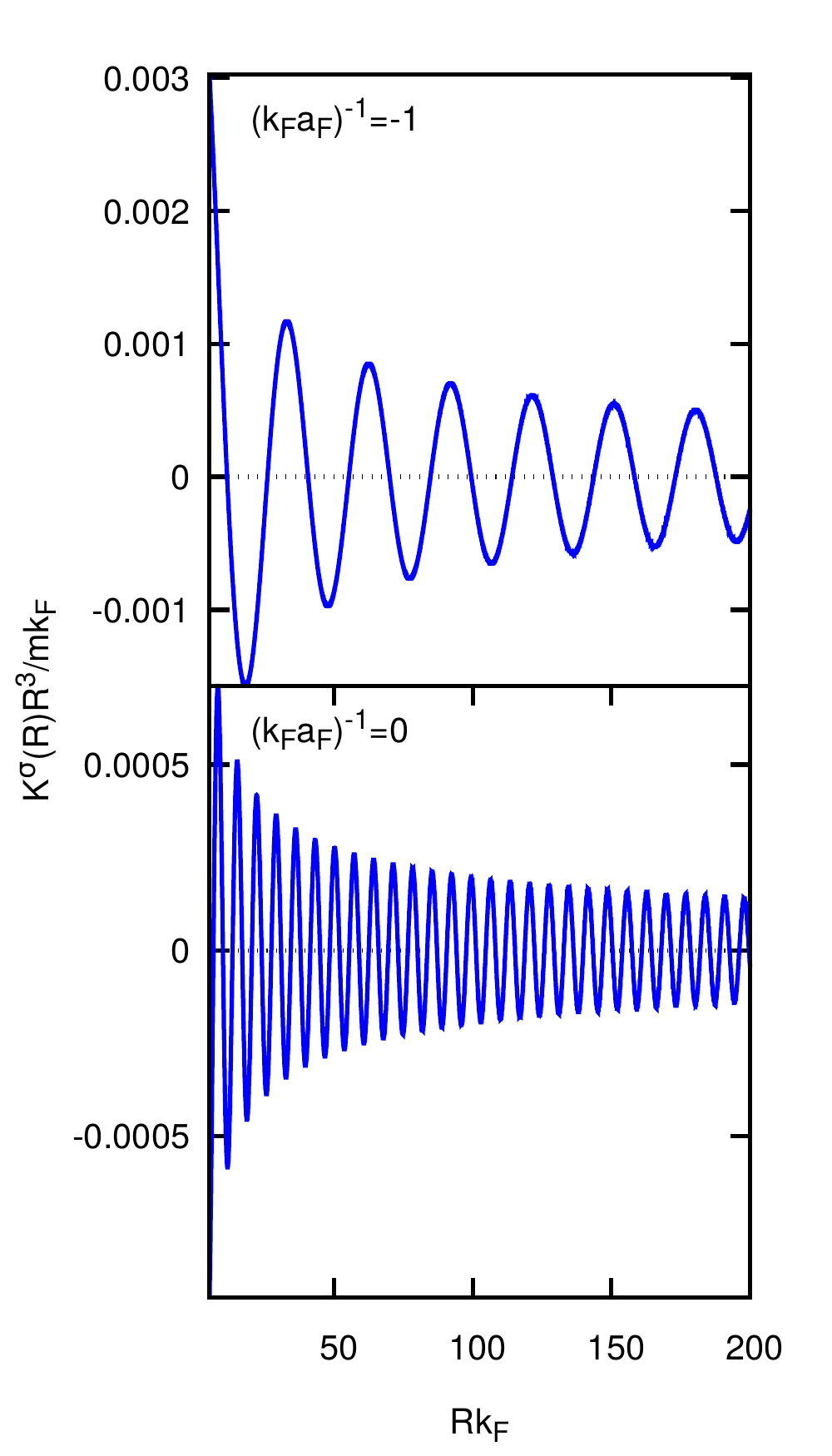}
\caption{(Color online) Radial profile of the kernel $K^{\sigma}(R)$ with $\sigma=10 k_{F}$ for two characteristic couplings at $T=0$. Here, $K^{\sigma}(R)$ has been multiplied by $R^{3}$
                                     to identify the exponent of the power-law behaviour of the tail.}
\label{Figure-5}
\end{center}
\end{figure}

\vspace{0.2cm}
\noindent
{\bf Spatial oscillations on the BCS side at zero temperature and the effect of temperature.}
Figure~\ref{Figure-4} shows the spatial profile of the kernel $K^{\sigma}(R)$ (multiplied by $R^{4}$) for the same couplings and temperatures of Fig.~\ref{Figure-2} and with a common value of $\sigma$.
Note that, at $T=0$ and for the couplings $(k_{F} a_{F})^{-1} =(-1.0,0.0)$, the kernel $K^{\sigma}(R)$ presents regular oscillations of wave vector $2 \, Q_{c}$, which get quickly damped as soon as the temperature is increased.
No oscillations are instead present at any temperature for the coupling $(k_{F} a_{F})^{-1} = 1.0$, as expected.

\vspace{0.2cm}
\noindent
{\bf Long-range tail of the kernel $\mathbf{K^{\sigma}(R)}$.}
To identify the decay rate of the amplitude of the oscillations of the kernel $K^{\sigma}(R)$ at large $R$, Fig.~\ref{Figure-5} shows $K^{\sigma}(R)$ 
multiplied by $R^{3}$ over an extended range of $R$ for two couplings at $T=0$.
In this way, the amplitude of $K^{\sigma}(R)$ is found to behave asymptotically like $R^{-3}$ for both couplings.
In Appendix~\ref{sec:appendix-A}, by studying a model function in $Q$-space whose Fourier transform in $R$-space can be evaluated analytically, we will verify that the $R^{-3}$ tail 
of the kernel $K^{\sigma}(R)$ stems from the fact that, in three dimensions, the kink singularity of the kernel $K(Q)$ extends over a sphere of finite radius.
Consistently, for the asymptotic kernel $K_{\infty}(Q)$ for which the kink singularity reduces to the single point $Q=0$, the power-law dependence of the tail 
of its Fourier transform $K^{\sigma}_{\infty}(R)$ becomes $R^{-4}$ and no oscillation occurs in this case (cf. also Eq.~(\ref{K-infty-R}) below).

\begin{figure}[t]
\begin{center}
\includegraphics[width=7.0cm,angle=0]{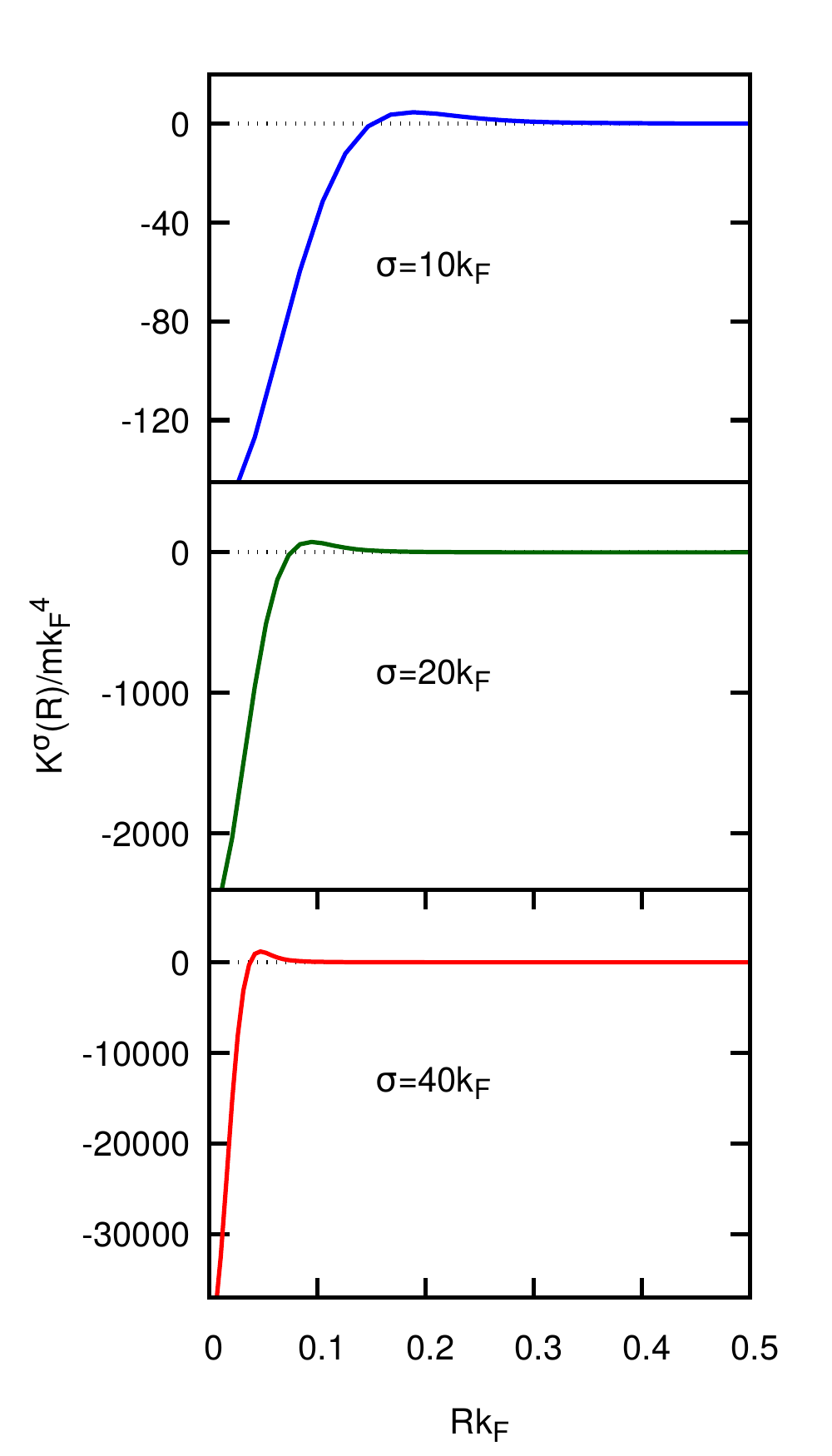}
\caption{(Color online) The behaviour of the kernel $K^{\sigma}(R)$ near the origin is shown for various values of $\sigma$, when $T=0$ and $(k_{F} a_{F})^{-1}=0$.}
\label{Figure-6}
\end{center}
\end{figure}

\vspace{0.2cm}
\noindent
{\bf Behaviour of $\mathbf{K^{\sigma}(R)}$ near the origin.}
The behaviour of the kernel $K^{\sigma}(R)$ near the origin depends on the large-$Q$ behaviour of the kernel $K^{\sigma}(Q)$.
This, in turn, depends on $\sigma$ but \emph{not} on coupling and temperature, as it can be seen from the expressions (\ref{large_Q-remaining_part}) and (\ref{kernel-new-definition}).  
Figure~\ref{Figure-6}   shows typical profiles of the kernel $K^{\sigma}(R)$ in the restricted range $R \, k_{F} \le 0.5$, for $T=0$, $(k_{F}a_{F})^{-1}=0$, and various values of $\sigma$. 
One notes from Fig.~\ref{Figure-6} that, by increasing the value of $\sigma$ from one panel to the next by a factor of two, a large increase results in the value 
of $K^{\sigma}(R)$ when $R \rightarrow 0$ (which eventually leads to a divergence when $\sigma \rightarrow \infty$). 
One can also verify that the zero of $K^{\sigma}(R)$ closest to the origin occurs at $R = R^{*} \simeq 1.52 / \sigma$, thereby approaching $R=0$ when $\sigma \rightarrow \infty$.
We have also verified that, in the restricted spatial range of Fig.~\ref{Figure-6}, plots with given $\sigma$ but different values of $T$ and $(k_{F}a_{F})^{-1}$ can hardly be distinguished from each other.

\begin{figure}[t]
\begin{center}
\includegraphics[width=6.5cm,angle=0]{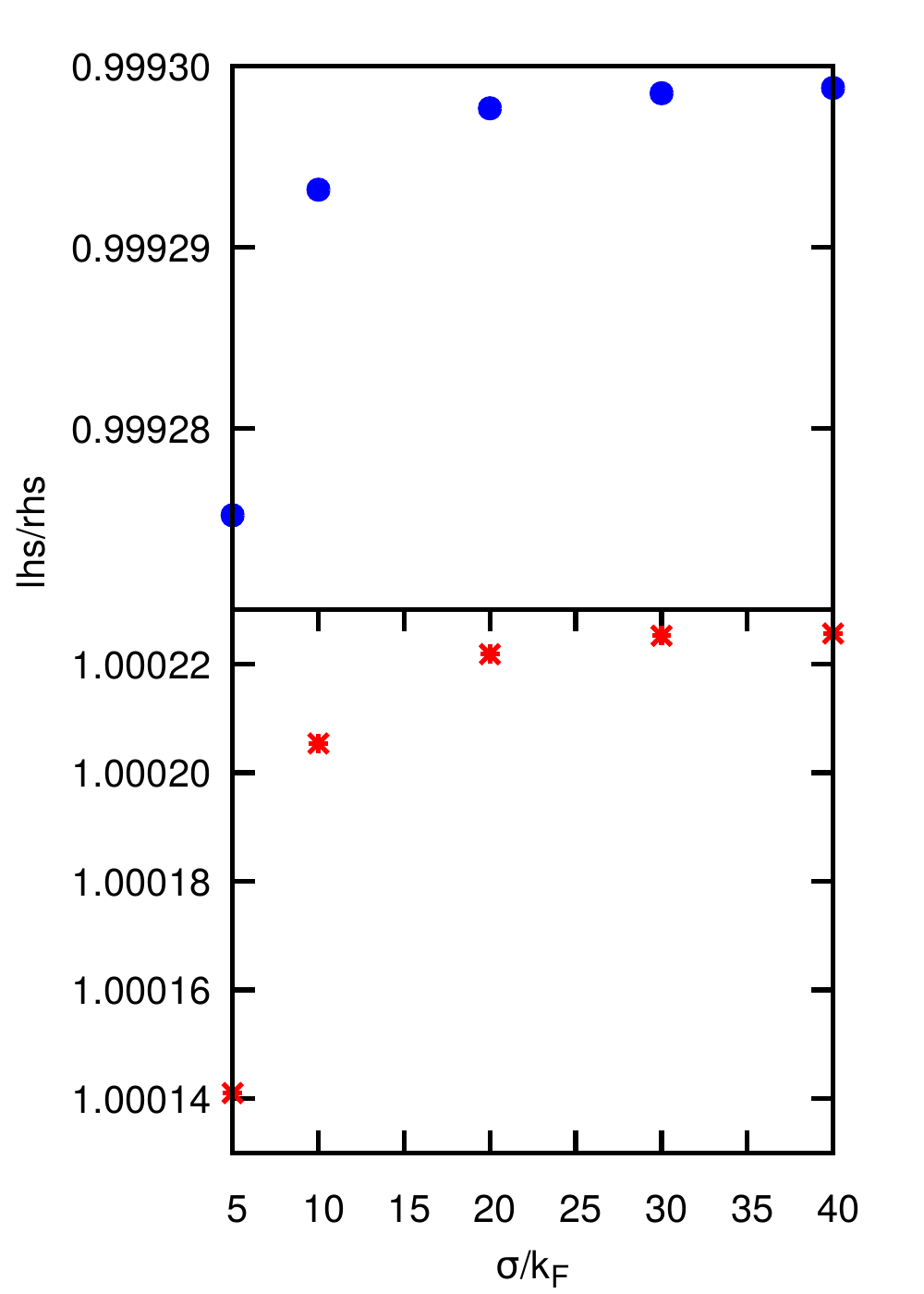}
\caption{(Color online) Ratio of the left-hand side (lhs) to the right-hand side (rhs) of both Eq.~(\ref{useful-identitiy-1}) (upper panel - dots) and Eq.~(\ref{useful-identitiy-2}) 
                                    (lower panel - stars) calculated for different values of $\sigma$, when $T=0$ and $(k_{F} a_{F})^{-1} = -1$.}
\label{Figure-7}
\end{center}
\end{figure}

\vspace{0.2cm}
\noindent
{\bf Numerical checks on the overall shape of $\mathbf{K^{\sigma}(R)}$.}
The following identities hold for the kernel $K^{\sigma}(R)$ at any coupling and temperature below $T_{c}$:
\begin{small}
\begin{eqnarray}
& & 4 \pi \! \int_{0}^{\infty} \!\! dR \, R^{2} \, K^{\sigma}(R) \, = \, K(\mathbf{Q}=0) \, = \, \mathcal{I}_{0} \, ,
\label{useful-identitiy-1} \\
& & 16 \pi \! \int_{0}^{\infty} \!\! dR \, R^{4} \, K^{\sigma}(R) \, = \, - \, \nabla^{2} \! \left(  K(\mathbf{Q}) \, e^{-\mathbf{Q}^{2}/\sigma^{2}} \right)_{\mathbf{Q}=0} 
\nonumber \\
& = & - \, \nabla^{2} \! \left. K(\mathbf{Q}) \right|_{\mathbf{Q}=0} + \frac{6}{\sigma^{2}} \, K(\mathbf{Q}=0) \, =  \, \frac{ 6 \, \mathcal{I}_{1}}{m} + \, \frac{ 6 \, \mathcal{I}_{0}}{\sigma^{2}} \, ,
\label{useful-identitiy-2} 
\end{eqnarray}
\end{small}

\noindent
where the notation of Eqs.~(\ref{non-local-LPDA-equation-uniform}) and (\ref{second-derivative-Q}) has been used.
These identities can be used as a check on the overall shape of the kernel $K^{\sigma}(R)$ obtained numerical for a given value of $\sigma$.
These checks are important especially on the BCS side of unitarity at $T=0$, where the kernel $K^{\sigma}(R)$ has rapid oscillations with a slowly decaying amplitude.
In this context, special care requires the $R$-integral on the left-hand side of Eq.~(\ref{useful-identitiy-2}), which is understood to contain also a Gaussian weight of the form 
$e^{-R^{2}/\sigma_{\mathrm{R}}^{2}}$ (such that $\sigma_{\mathrm{R}} \rightarrow \infty$ at the end of the calculation).
Otherwise, this integral would not converge owing the $1/R^{3}$ tail of $K^{\sigma}(R)$.
Accordingly, we shall interpret the integral
\begin{equation}
\int \! \frac{d \mathbf{R}}{\pi^{3}} \,\, e^{2 i \mathbf{Q} \cdot \mathbf{R}} \, e^{-R^{2}/\sigma_{\mathrm{R}}^{2}} = \left( \frac{\sigma_{\mathrm{R}}}{\sqrt{\pi}} \right)^{3}
e^{-\mathbf{Q}^{2} \sigma_{\mathrm{R}}^{2}} \,\,\, \longrightarrow \,\,\, \delta(\mathbf{Q})
\label{Gaussian-delta-function}
\end{equation}
\noindent
as approaching the Dirac delta function $\delta(\mathbf{Q})$ in the sense of distributions when $\sigma_{\mathrm{R}} \rightarrow \infty$ \cite{Lighthill-1958}.

Figure~\ref{Figure-7} shows the ratio of the left-hand side (lhs) to the right-hand side (rhs) of both Eq.~(\ref{useful-identitiy-1}) (upper panel) and Eq.~(\ref{useful-identitiy-2}) (lower panel)
for different values of $\sigma$ when $T=0$ and $(k_{F} a_{F})^{-1} = -1$ 
(for this calculation the value $\sigma_{\mathrm{R}} = 50 k_{F}^{-1}$ has been used in accordance with the above argument).
The steady convergence of these results gives us confidence about the stability of our numerical calculations of the kernel $K^{\sigma}(R)$ for increasing $\sigma$ (notwithstanding the divergence of this kernel at $R=0$ for increasing $\sigma$ - cf. Eq.~(\ref{K-infty-small-R}) below).
In this context, we have also verified numerically that, for growing $\sigma$, the profile of $K^{\sigma}(R)$ tends uniformly toward an asymptotic profile, with a convergence rate that becomes slower as $R$ gets close to $R=0$ (where the $\sigma$-dependent singular behaviour shown in Fig.~\ref{Figure-6} occurs).

\vspace{-0.5cm}
\begin{figure}[h]
\begin{center}
\includegraphics[width=7.8cm,angle=0]{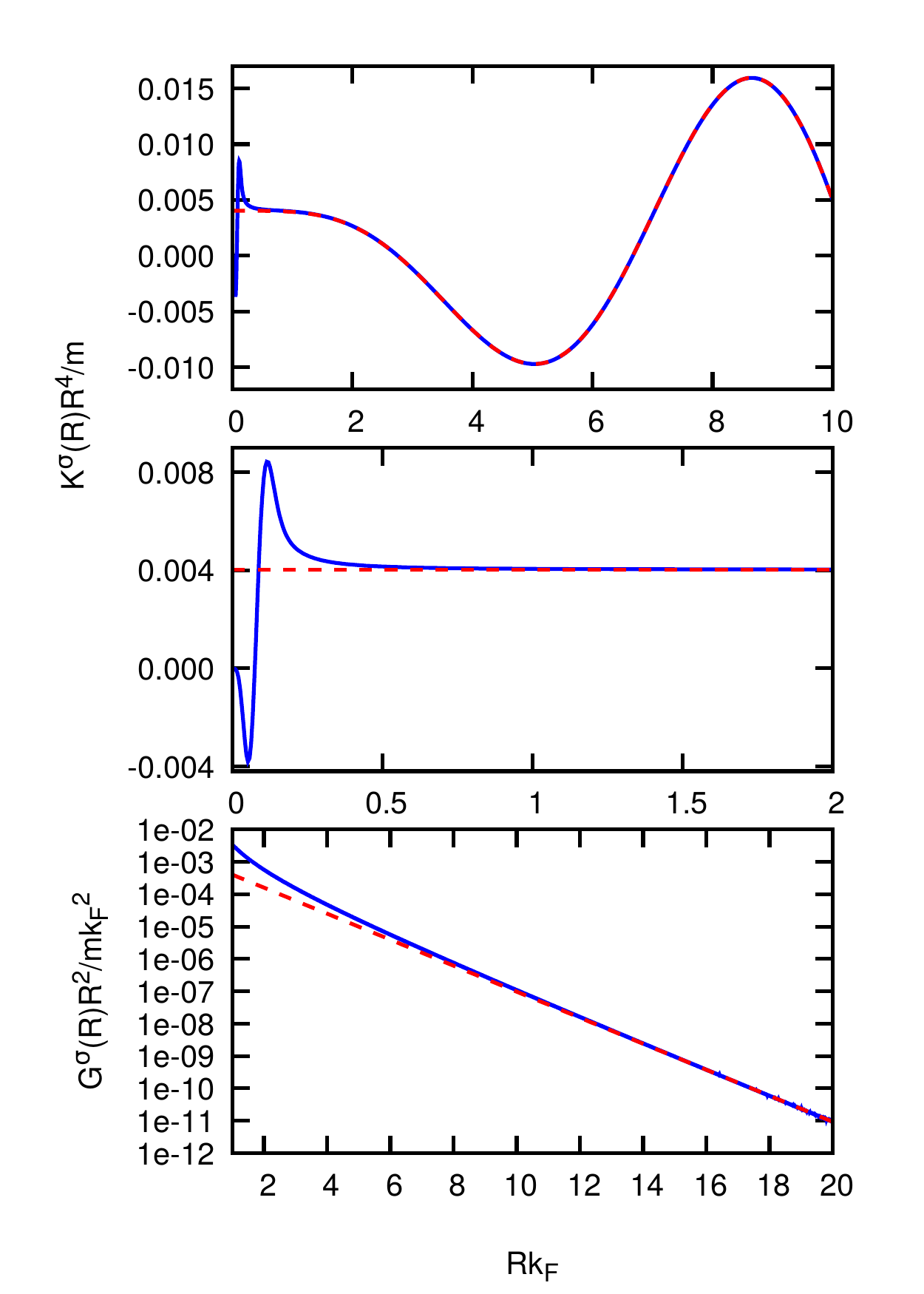}
\caption{(Color online)  \emph{Upper panel}: Model function $K_{\mathrm{model}}^{\sigma}(R)$ multiplied by $R^{4}$ vs $R$, for $\alpha = m /(4 \pi)$ and $Q_{0} = 0.447 k_{F}$.
                                     The asymptotic expression (\ref{K-model-large-R}) (dashed line) is compared with the full expression given by 
                                     Eqs. (\ref{FT-K-model-1}), (\ref{integral-J1-solution}), and (\ref{integral-J2-solution}) (full line). 
                                     \emph{Middle panel}: Asymptotic kernel $K_{\infty}^{\sigma}(R)$ multiplied by $R^{4}$ vs $R$. 
                                     The asymptotic expression (\ref{K-infty-large-R}) (dashed line) is compared with the full expression given by Eq.~(\ref{K-infty-R}) (full line). 
                                     \emph{Lower panel}: Model function $G^{\sigma}(R)$ multiplied by $R^{2}$ vs $R$, obtained numerically from Eq.~(\ref{smooth-model-function})
                                     with $\alpha = m /(4 \pi)$ and $P_{0} = 0.447 k_{F}$ (full line). The dashed line represents the asymptotic behavior $A \, \exp\{ - R/L \}$ with
                                     $A = 1.02 \times 10^{-3}$ and $L P_{0}= 0.483$ (which is $3 \%$ off the expected value $L P_{0}= 0.5$). In all cases, $\sigma = 20 k_{F}$.}
\label{Figure-8}
\end{center}
\end{figure}

\vspace{0.6cm}
\noindent
{\bf The kernel $\mathbf{K_{\infty}^{\sigma}(R)}$.}
The asymptotic kernel (\ref{asymptotic-kernel}) in $\mathbf{Q}$-space is a particular case of the model function studied analytically in Appendix A, where one sets $\alpha = m / (4 \pi)$ and $Q_{0} = 0$ in Eq.~(\ref{model-function-Q-space}) therein.
The analysis carried out in Appendix A for the Fourier transform $K_{\mathrm{model}}^{\sigma}(R)$ then yields for the function $K_{\infty}^{\sigma}(R)$ in the limit $Q_{0} \rightarrow 0$:
\begin{eqnarray}
& & K_{\infty}^{\sigma}(R) = \frac{ i \, m \sqrt{\pi} \sigma^{3}}{8 \pi^{3} R} 
\nonumber \\
& \times & \frac{d^{2}}{d y^{2}} \left[ e^{-\frac{y^{2}}{4}} \left( \! \mathrm{erfc} \left(\! i \frac{y}{2} \right) - \mathrm{erfc} \left(\! -i \frac{y}{2} \right) \right) \right] _{y=2 \sigma R}
\label{K-infty-R}
\end{eqnarray}

\noindent
for all values of $R$.
Knowledge of the overall profile of $K_{\infty}^{\sigma}(R)$ given by Eq.~(\ref{K-infty-R}) will be useful in what follows.

The profile of the kernel $K_{\infty}^{\sigma}(R)$ is shown in Fig.~\ref{Figure-8} (middle panel) over an extended range of $R$.
For comparison, the profile of the model function $K_{\mathrm{model}}^{\sigma}(R)$ studied analytically in Appendix A (from which $K_{\infty}^{\sigma}(R)$ is obtained 
in the limit $Q_{0} \rightarrow 0$) is also shown in Fig.~\ref{Figure-8} (upper panel), when $Q_{0}$ equals the critical value of $Q_{c}$ at unitarity
taken from Fig.~\ref{Figure-1}.
In both panels, $\sigma = 20 k_{F}$.

Knowledge of the limiting behaviours of $K_{\infty}^{\sigma}(R)$ will also be useful in what follows.
When $R \rightarrow \infty$ we obtain from Eq.~(\ref{K-model-large-R}) with $\alpha = m / (4 \pi)$ and $Q_{0} = 0$:
\begin{equation}
K_{\infty}^{\sigma}(R \rightarrow \infty) \, \simeq \, \frac{m}{8 \, \pi^{3} \, R^{4}}
\label{K-infty-large-R}
\end{equation}
\noindent
which shows no oscillatory behaviour, contrary to the expression (\ref{K-model-large-R}) with $Q_{0} \ne 0$.
When $R \rightarrow 0$ we obtain from Eq.~(\ref{K-model-small-R}), again with $\alpha = m / (4 \pi)$ and $Q_{0} = 0$:
\begin{equation}
K_{\infty}^{\sigma}(R \rightarrow 0) \, \simeq - \, \frac{m \, \sigma^{4}}{2 \, \pi^{3}}
\label{K-infty-small-R}
\end{equation}
\noindent
which, apart from a minor correction, diverges like the expression (\ref{K-model-small-R}) in the limit $\sigma \rightarrow \infty$.

\vspace{0.2cm}
\noindent
{\bf Recovering the exponential behaviour of the kernel $\mathbf{K^{\sigma}(R)}$.}
It is apparent from Fig.~\ref{Figure-2} that the kink singularity of the kernel $K(Q)$, which occurs at $T=0$ on the weak-coupling (BCS) side of the crossover, disappears \emph{either} by increasing the temperature toward $T_{c}$ \emph{or} by moving to the BEC side of the crossover even at $T=0$. 
In both cases, the kernel $K(Q)$ becomes a smooth function of $Q$.
To mimic this behaviour, we consider the following simple model function:
\begin{equation}
G(Q) = - \, \alpha \, \sqrt{ P_{0}^{2} + Q^{2} }
\label{smooth-model-function}
\end{equation}
\noindent
with $\alpha = m /(4 \pi)$, which (apart from an overall constant shift) has the same kind of small-$Q$ and large-$Q$ behaviour of the full kernel $K(Q)$.
We then multiply this function by  $e^{-\mathbf{Q}^{2} / \sigma^{2}}$ as it was done in Eqs.~(\ref{kernel-new-definition}) and (\ref{asymptotic-kernel-new-definition}), 
and calculate numerically the spatial Fourier transform of the ensuing function $G^{\sigma}(Q)$ in three dimensions.
The result is shown in Fig.~\ref{Figure-8} (lower panel ) for $\sigma = 20 k_{F}$ and $P_{0}$ equal to the value of $Q_{c}=0.447 k_{F}$ taken from Fig.~\ref{Figure-1} at unitarity.
From this plot one concludes that the large-$R$ behaviour of the function $G^{\sigma}(R)$ has the form $ A \exp\{-R/L\} / R^{2}$ where the length $L$ is of the order of $(2 P_{0})^{-1}$. 
It is interesting to note that this behaviour coincides with that of the kernel yielding the linear terms in the Gor'kov derivation of the Ginzburg-Landau equation \cite{Gorkov-1959}, which is valid on the BCS side of the crossover and close to $T_{c}$ \cite{FW-1971}.

\begin{figure}[t]
\begin{center}
\includegraphics[width=6.5cm,angle=0]{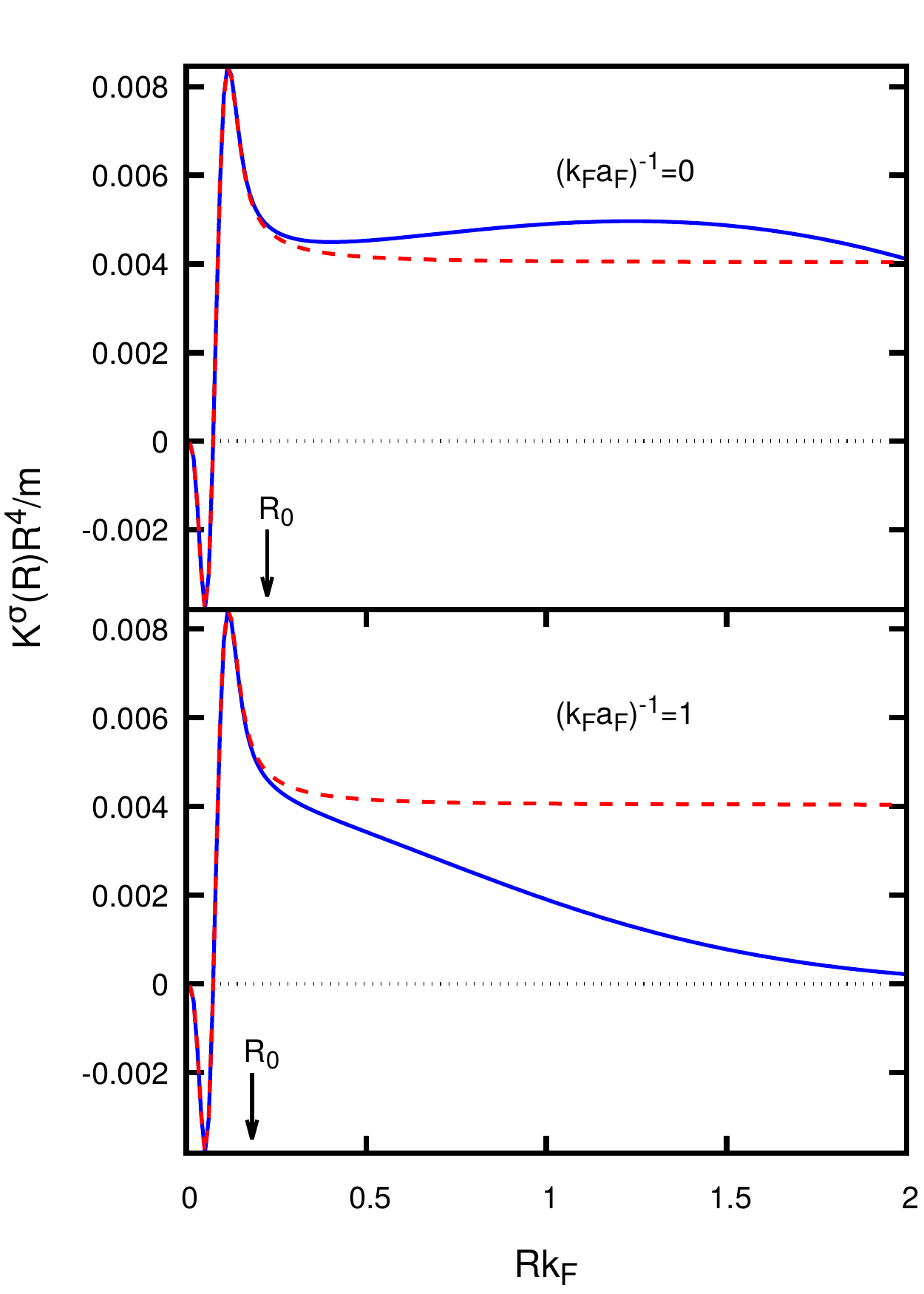}
\caption{(Color online) The kernels $K^{\sigma}(R)$ (full line) and $K^{\sigma}_{\infty}(R)$ (dashed line) are shown near the origin for $(k_{F} a_{F})^{-1}=0$ (upper panel) 
                                     and $(k_{F} a_{F})^{-1}=1.0$ (lower panel). In both cases, $T=0$ and $\sigma = 20 k_{F}$. The kernels are multiplied by $R^{4}$, making thus easier 
                                     to visualize the point $R_{0}$ (identified by a vertical arrow in each panel) at which they start to deviate from each other within $2 \%$.}
\label{Figure-9}
\end{center}
\end{figure}

These results imply that when passing, from a function like that given by Eq.~(\ref{model-function-Q-space}) with a kink singularity on the real $Q$-axis, to a function like 
that given by Eq.~(\ref{smooth-model-function}) where the branch-cut singularity resides in the complex plane at $\pm i P_{0}$, the large-$R$ behaviour of the corresponding Fourier transforms exhibits a drastic change, from the $\sin(2 Q_{0} R) / R^{3}$ behaviour of Eq.~(\ref{K-model-large-R}) to the $\exp\{-R/L\} / R^{2}$ behaviour of $G^{\sigma}(R)$.

\vspace{0.05cm}
\begin{center}
{\bf C. The regularized kernel $\mathbf{K_{\mathrm{reg}}^{\sigma}(R)}$}
\end{center}
\vspace{-0.2cm}

From Eqs.~(\ref{K-infty-small-R}) and (\ref{K-model-small-R}) the kernels $K^{\sigma}(R)$ and $K_{\infty}^{\sigma}(R)$ are seen to tend to a common value when $R \rightarrow 0$, 
provided $\sigma$ is large enough.
We have verified numerically that this result remains true over a finite (albeit small) range of $R$, where $K^{\sigma}(R)$ and $K_{\infty}^{\sigma}(R)$ are seen to essentially coincide with each other.
An example is shown in Fig.~\ref{Figure-9} for a given value of $\sigma$ when $T=0$ and $(k_{F} a_{F})^{-1} = (0.0,1.0)$. 
Since the small-$R$ profile common to $K^{\sigma}(R)$ and $K_{\infty}^{\sigma}(R)$ is associated with the common large-$Q$ behaviour of the corresponding kernels in $Q$-space, which is independent of coupling and temperature, the results of Fig.~\ref{Figure-9} suggest us to adopt the following procedure which allows us to concentrate on the large-$R$ behaviour of the kernel $K^{\sigma}(R)$, that instead depends on both coupling and temperature.

\begin{figure}[t]
\begin{center}
\includegraphics[width=6.5cm,angle=0]{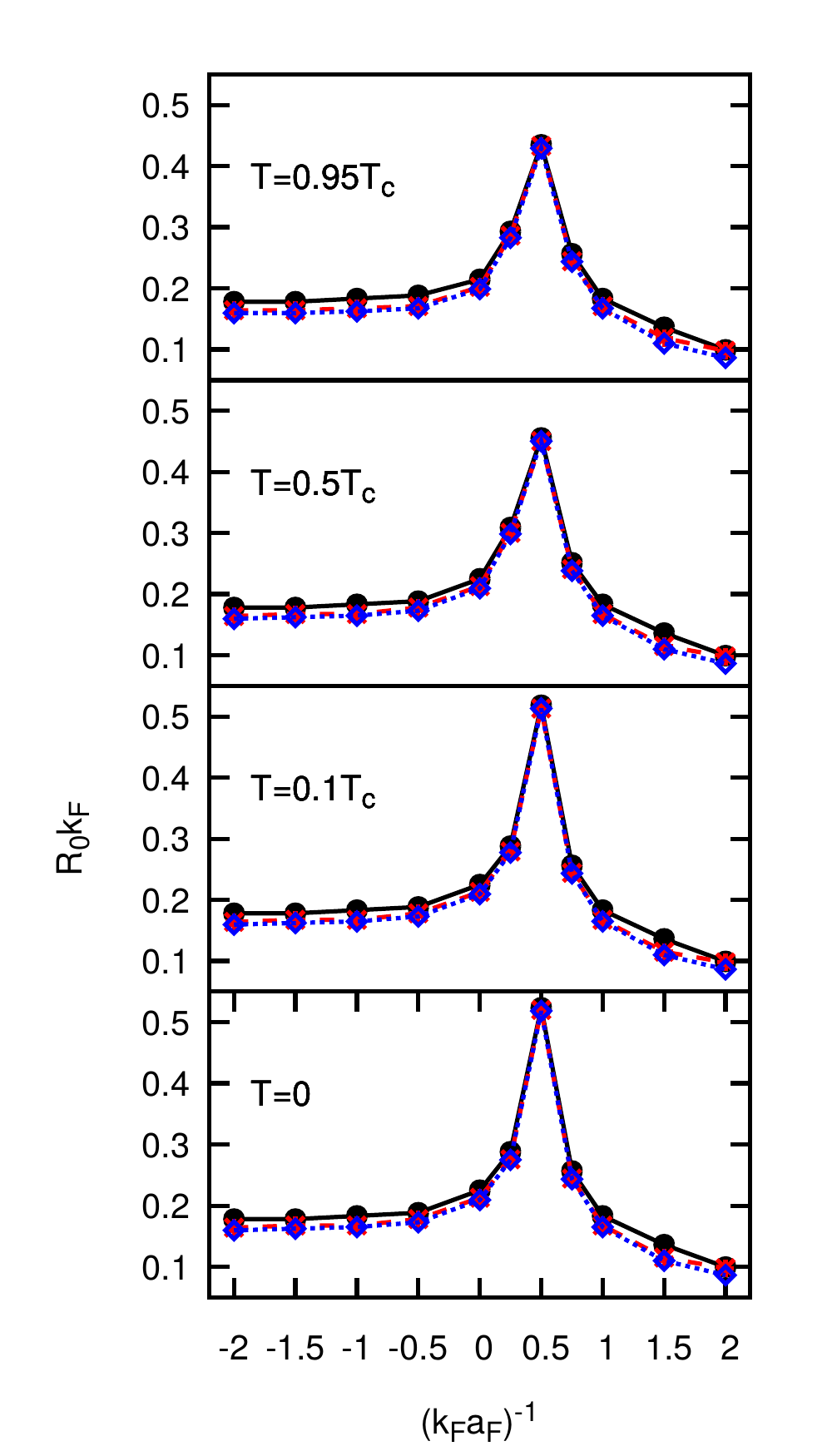}
\caption{(Color online) Coupling dependence of the point $R_{0}$ at which the two kernels $K^{\sigma}(R)$ and $K^{\sigma}_{\infty}(R)$ start to deviate from each other within $2 \%$,  
                                    for several temperatures. Here, $\sigma = 20 k_{F}$ (dots), $\sigma = 30 k_{F}$ (stars), and $\sigma = 40 k_{F}$ (diamonds).}
\label{Figure-10}
\end{center}
\end{figure}

For given coupling and temperature, we move from $R=0$ outwards and search for the point $R_{0}$ at which the two kernels $K^{\sigma}(R)$ and $K^{\sigma}_{\infty}(R)$ start to deviate from each other, 
say, within $2 \%$.
This can be done for a set of values of $\sigma$, thus monitoring the convergence of the results for $\sigma \rightarrow \infty$.
The results of this procedure are shown in Fig.~\ref{Figure-10} throughout the BCS-BEC crossover, for several temperatures and three different 
values of $\sigma$. 
In all cases, the values of $R_{0}$ are not larger than $1/(2 k_{F})$, which represents a small length scale compared with the overall spatial extent 
of the kernel $K^{\sigma}(R)$ \cite{footnote-1}. 
Note also that $R_{0}$ depends weakly on $\sigma$ for all couplings and temperatures.
This result is remarkable, in light of the fact that in the small-$R$ region (compared to $k_{F}^{-1}$) both kernels $K^{\sigma}(R)$ and 
$K_{\infty}^{\sigma}(R)$ change instead considerably by varying $\sigma$.
For instance, from Fig.~\ref{Figure-6} the positions of both the first zero and the first maximum in $K^{\sigma}(R)$ are seen to decrease by a factor of two from $\sigma = 20 k_{F}$ to $\sigma = 40 k_{F}$, while from Fig.~\ref{Figure-10} $R_{0}$ is seen corrispondigly to change only by a few percents.

Once the point $R_{0}$ is identified by this procedure, for given $\sigma$ we define a \emph{regularized kernel} $K_{\mathrm{reg}}^{\sigma}(R)$ as follows:
\begin{equation}
K_{\mathrm{reg}}^{\sigma}(R) = \left\{ \begin{array}{cc}       0              &   \,\,\, (R < R_{0})    \\
                                                                                     K^{\sigma}(R)   &   \,\,\, (R \ge R_{0})       \end{array}  \right. \, .
\label{regularized-kernel-R-space}
\end{equation}
\noindent
In this way, a ``hole'' about $R=0$ is effectively introduced in the original kernel $K^{\sigma}(R)$, thereby avoiding its strong divergence for $\sigma \rightarrow \infty$ 
but at the same time not affecting the determination of its spatial range, for which the behaviour when $R \le R_{0}$ is irrelevant.

\vspace{0.05cm}
\begin{center}
{\bf D. Spatial range of the kernel $\mathbf{K(R)}$ as a function of coupling and temperature}
\end{center}
\vspace{-0.2cm}

To determine the spatial range $\xi_{\mathrm{K}}^{\sigma}$ of the kernel $K^{\sigma}(R)$ (through its regularized version $K_{\mathrm{reg}}^{\sigma}(R)$) as a function of coupling and temperature, it is convenient to distinguish two cases when the kernel $K^{\sigma}(R)$ for \emph{large} $R$ has:

\noindent
(i) An oscillatory behaviour, like at $T=0$ from the BCS to the unitary regime;

\noindent
(ii) An exponential behaviour, like at $T=0$ in the BEC regime or when approaching $T_{c}$ even in the BCS and unitary regimes.

\begin{figure}[t]
\begin{center}
\includegraphics[width=7.6cm,angle=0]{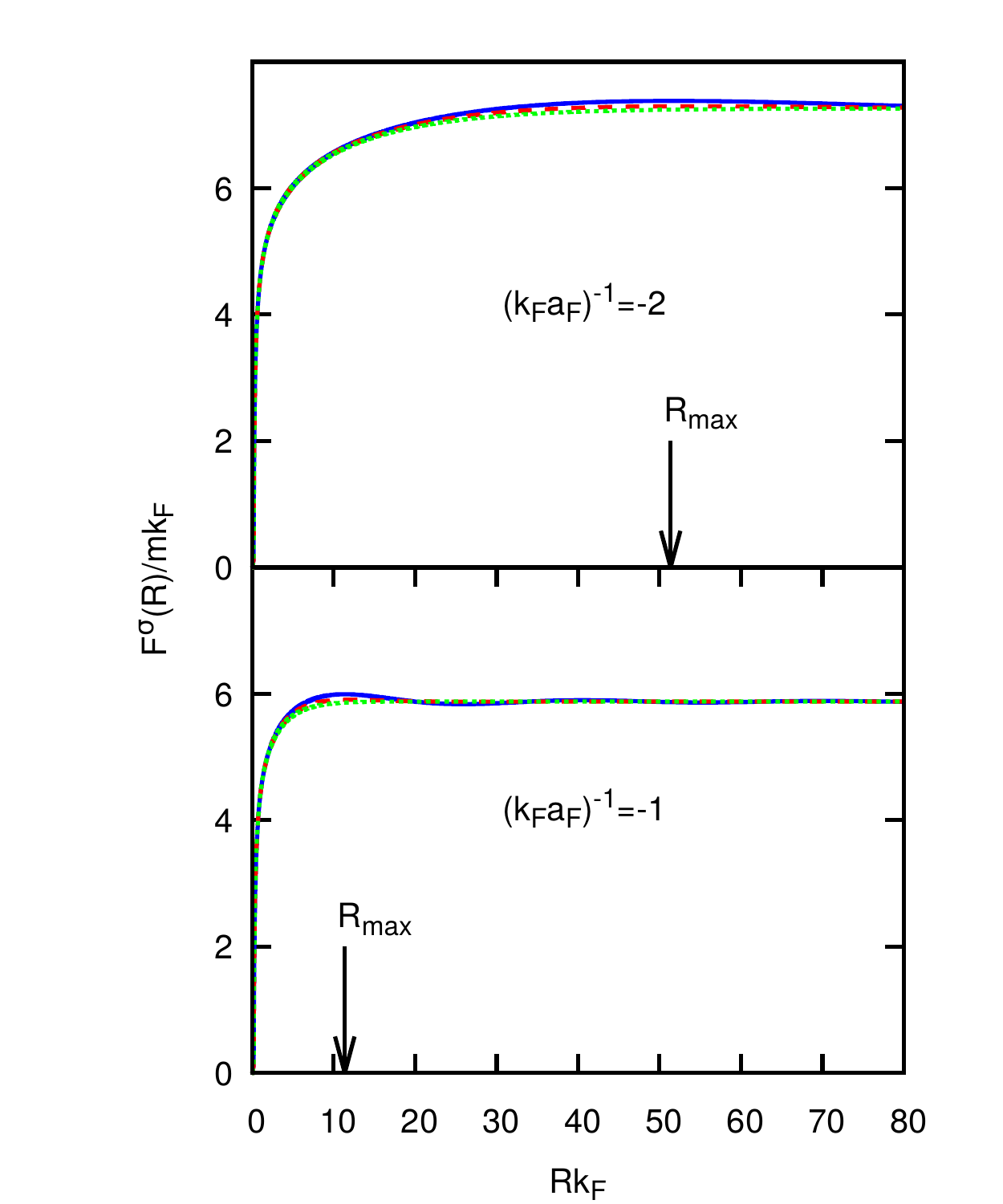}
\caption{(Color online) The function $F^{\sigma}(R)$ given by Eq.~(\ref{function-F}) with $\sigma = 20 k_{F}$ is shown vs $R$, when $(k_{F} a_{F})^{-1}= -2.0$ (upper panel) and 
                                     $(k_{F} a_{F})^{-1}= -1.0$ (lower panel), and for temperatures: $T=0$ (full line); $T=0.5 T_{c}$ (dashed line); $T=0.95 T_{c}$ (dotted line). 
                                     In both panels, the arrows identify the value $R_{\mathrm{max}}$ at which $F^{\sigma}(R)$ has reached its first maximum.}
\label{Figure-11}
\end{center}
\end{figure}

For case (i), the spatial range $\xi_{\mathrm{K}}^{\sigma}$ of the kernel $K^{\sigma}(R)$ is determined by considering the behaviour of the function:
\begin{equation}
F^{\sigma}(R) = \!\! \int_{R_{0}}^{R} \! dR' \, R'^{2} \, K^{\sigma}(R') = \!\! \int_{0}^{R} \! dR' \, R'^{2} \, K_{\mathrm{reg}}^{\sigma}(R') \, .
\label{function-F}
\end{equation}
\noindent
For given $\sigma$, this function converges asymptotically to a finite value $F^{\sigma}(\infty)$ in the limit $R \rightarrow \infty$, and this is so even at $T=0$ when its integrand $K^{\sigma}(R)$ has characteristic oscillations of wave vector $2 Q_{0}$ with amplitude decaying like $1/R^{3}$ for large $R$.
Typical examples of the behaviour of $F^{\sigma}(R)$ vs $R$ are shown in Fig.~\ref{Figure-11} for $\sigma = 20 k_{F}$, $(k_{F} a_{F})^{-1} = (-1.0,-2.0)$, and
$T = (0.0,0.5,0.95) T_{c}$.
At $T=0$, one sees that  $F^{\sigma}(R)$ has essentially converged to its asymptotic value $F^{\sigma}(\infty)$ as soon as it reaches the first maximum at
$R_{\mathrm{max}}$, past which $F^{\sigma}(R)$ shows only a minor oscillatory behaviour around $F^{\sigma}(\infty)$.
At $T = 0.5 T_{c}$ and $T = 0.95 T_{c}$, on the other hand, $F^{\sigma}(R)$ shows no oscillatory behaviour and reaches monotonically the asymptotic value $F^{\sigma}(\infty)$ at about the same value of $R_{\mathrm{max}}$ identified at $T=0$. 
This implies that, apart from the presence or absence of minor oscillations, for both couplings the overall shape of $F^{\sigma}(R)$ remains essentially the same upon varying the temperature. 
Accordingly, for coupling values $(k_{F} a_{F})^{-1} \lesssim 0.50$ we identify the value of $R_{\mathrm{max}}$ obtained by the above procedure with the spatial range $\xi_{\mathrm{K}}^{\sigma}$ of the kernel $K^{\sigma}(R)$ at $T=0$ (expecting further that this range should only slightly depend on temperature).

\begin{figure}[t]
\begin{center}
\includegraphics[width=8.8cm,angle=0]{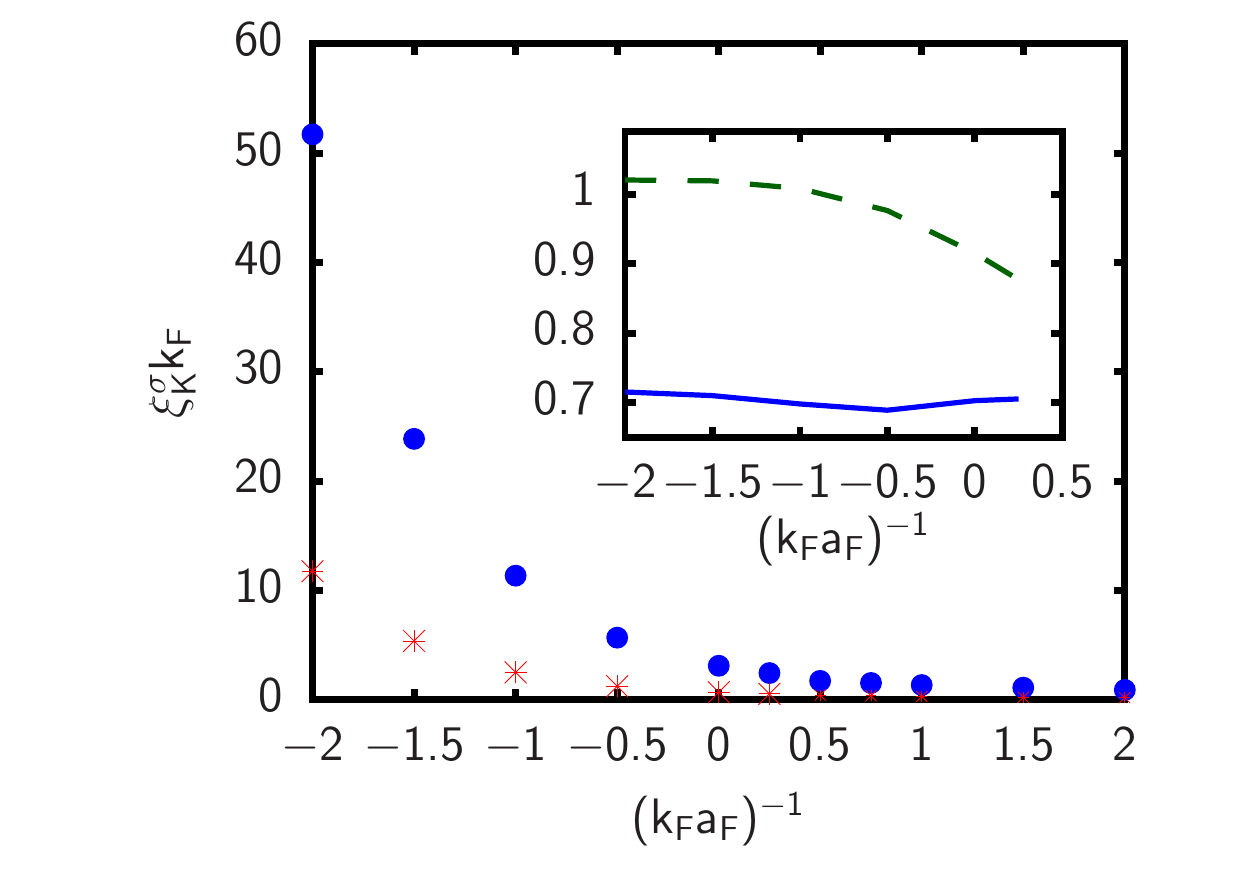}
\caption{(Color online) Coupling dependence of the spatial range of the kernel $K^{\sigma}(R)$.
                                    The values obtained for $\xi_{\mathrm{K}}^{\sigma}$ at $T=0$ (dots) are compared with the values obtained for $L^{\sigma}$ at $T = 0.99 T_{c}$ (stars)
                                    The inset shows the coupling dependence of the ratio $\pi L^{\sigma}(0.99 T_{c}) / \xi_{K}^{\sigma}(T=0)$ (full line) a well as of the quantity 
                                    $k_{F} / [2 \pi m k_{B} T_{c} L^{\sigma}(0.99 T_{c})]$ (dashed line). In all cases, $\sigma = 20 k_{F}$.}
\label{Figure-12}
\end{center}
\end{figure}

For case (ii), the product $R^{2} K^{\sigma}(R)$ is found to behave like $\exp\{-R/L^{\sigma}\}$ for large $R$, such that an exponential fit can be made directly on this product to extract the characteristic length $L^{\sigma}$.
We have performed this fit, at $T=0$ for $(k_{F} a_{F})^{-1} \gtrsim 0.50$ and at $T=0.99T_{c}$ across the whole BCS-BEC crossover, again with the value $\sigma = 20 k_{F}$.
To connect with continuity the values of $L^{\sigma}$ obtained here at $T=0$ for $(k_{F} a_{F})^{-1} \gtrsim 0.50$ with the values of $\xi_{\mathrm{K}}^{\sigma}$ obtained previously for 
$(k_{F} a_{F})^{-1} \lesssim 0.50$, we have set $L^{\sigma} = \gamma \xi_{\mathrm{K}}^{\sigma}$ and determined the constant $\gamma$ in such a way that $\xi_{\mathrm{K}}^{\sigma} \rightarrow \pi a_{F} / \sqrt{2}$ upon approaching the BEC limit. 
By carrying out these calculations up to $(k_{F} a_{F})^{-1} = 4.0$, we have obtained the value $\gamma \simeq 0.2 \simeq 2 / \pi^{2}$.

Figure~\ref{Figure-12} shows the coupling dependence throughout the whole BCS-BEC crossover of the range $\xi_{\mathrm{K}}^{\sigma}$ obtained at $T=0$ by the above ``mixed'' procedure (dots), together with the coupling dependence of the range $L^{\sigma}(0.99T_{c})$ obtained by the above exponential fit at $T = 0.99T_{c}$ (stars).
In addition, the inset of Fig.~\ref{Figure-12} combines these data in the coupling dependence of the ratio $\pi L^{\sigma}(0.99 T_{c}) / \xi_{K}^{\sigma}(T=0)$ (full line), which turns out to be about $0.7$ for all couplings in the range $(k_{F} a_{F})^{-1} \le 0.25$.
Such a weak temperature dependence from $T=0$ up to $T_{c}$, that we have obtained for the range of the kernel $K^{\sigma}(R)$ irrespective of coupling, is in line with the behaviour of the size of the Cooper pairs obtained in Ref.~\cite{Palestini-2014}.
The inset of Fig.~\ref{Figure-12} shows also the coupling dependence of the quantity $k_{F} / [2 \pi m k_{B} T_{c} L^{\sigma}(0.99 T_{c})]$ for $(k_{F} a_{F})^{-1} \le 0.25$ (dashed line), which in the BCS limit and close to $T_{c}$ is expected to equal unity according to an analytic result due to Gor'kov \cite{FW-1971}.
Remarkably, our numerical calculations approximately reproduce this result not only in the BCS limit but also across unitarity.

\begin{figure}[t]
\begin{center}
\includegraphics[width=9.0cm,angle=0]{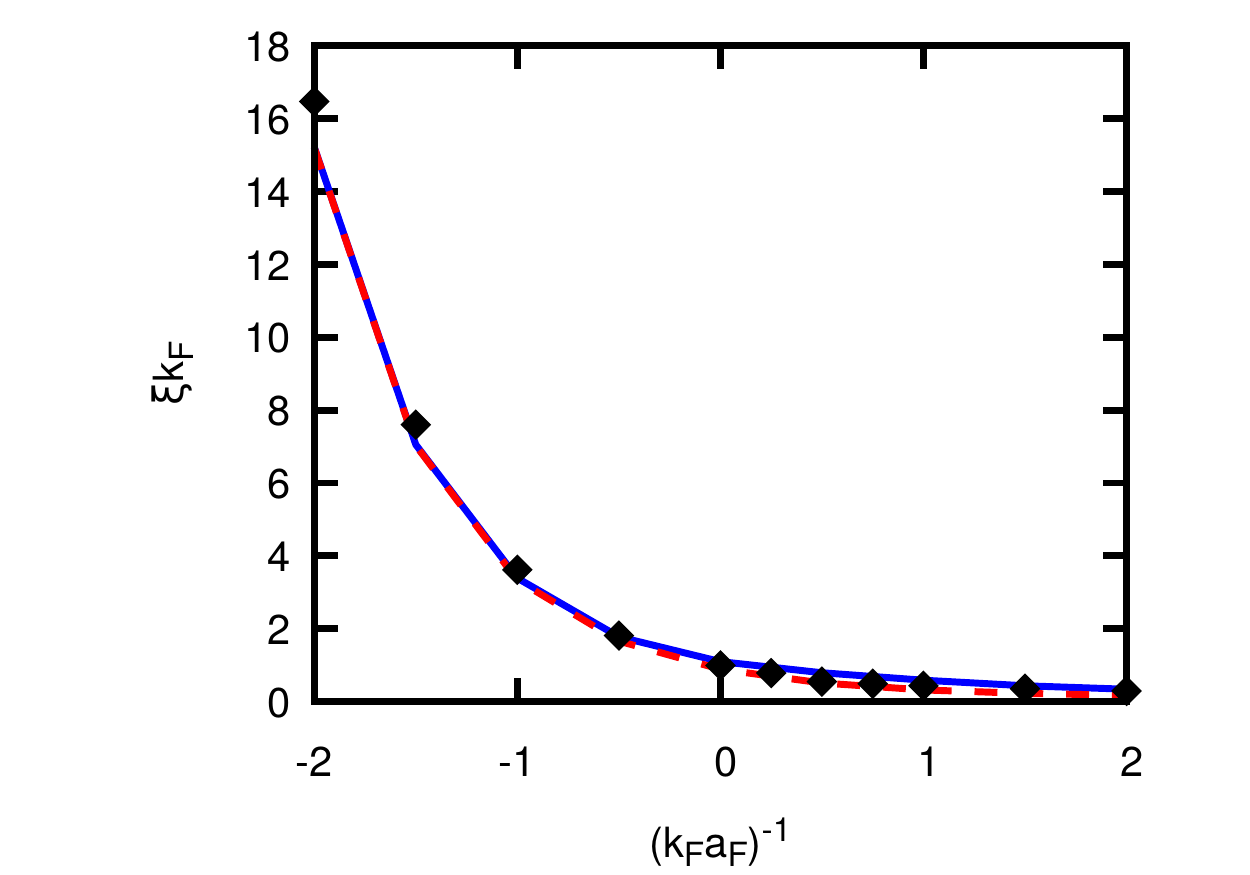}
\caption{(Color online) The coupling dependence of $\xi_{\mathrm{pair}}$ at $T=0$, as obtained in Ref.~\cite{Pistolesi-1994} (full line) and by the approximate expression 
                                     $( 2 \sqrt{2} Q_{c}^{\mathrm{L}} )^{-1}$ (dashed line) with $Q_{c}^{\mathrm{L}}$ taken from Fig.~\ref{Figure-1}, is compared with the coupling dependence 
                                     of $\xi_{\mathrm{K}}^{\sigma}/\pi$ for $\sigma = 20 k_{F}$ taken from the data of Fig.~\ref{Figure-12} (diamonds).}
\label{Figure-13}
\end{center}
\end{figure}

\vspace{0.05cm}
\begin{center}
{\bf E. Comparison of the length scales $\xi_{K}$ and $\xi_{\mathrm{pair}}$}
\end{center}
\vspace{-0.2cm}

The coupling behaviour of $\xi_{\mathrm{K}}^{\sigma}$ reported in Fig.~\ref{Figure-12} is reminiscent of the coupling behaviour of the pair coherence length $\xi_{\mathrm{pair}}$ throughout the BCS-BEC crossover, which was obtained originally at $T=0$ in Ref.~\cite{Pistolesi-1994} in terms of the pair correlation function of opposite-spin fermions at the mean-field level.
For the present purposes, it can be useful to relate $\xi_{\mathrm{pair}}$ at $T=0$ also with the approximate expression $( 2 \sqrt{2} Q_{c}^{\mathrm{L}} )^{-1}$, where $Q_{c}^{\mathrm{L}}$ is the Landau critical wave vector given by Eq.~(\ref{Q_c-Landau-criterion}).
In the weak-coupling (BCS) limit, this expression yields $( 2  \sqrt{2} Q_{c}^{\mathrm{L}} )^{-1}  \simeq k_{F} / ( 2 \sqrt{2} m \Delta)$, which coincides with the limiting value of $\xi_{\mathrm{pair}}$
obtained in Ref.~\cite{Pistolesi-1994} at $T=0$. 
In the strong-coupling (BEC) limit, on the other hand, $( 2 \sqrt{2} Q_{c}^{\mathrm{L}} )^{-1} \simeq a_{F}/(2 \sqrt{2})$ equals $\xi_{\mathrm{pair}}/2$ and vanishes in the relevant limit
$k_{F} a_{F} \ll 1$.
In addition, in the intermediate coupling region $-1 \lesssim (k_{F}\, a_{F})^{-1} \lesssim +1$ the expression $( 2 \sqrt{2} Q_{c}^{\mathrm{L}} )^{-1}$ approximates reasonably well 
the values of $\xi_{\mathrm{pair}}$ at $T=0$ obtained in Ref.~\cite{Pistolesi-1994}.

Figure~\ref{Figure-13} compares the coupling dependence of $\xi_{\mathrm{pair}}$ at $T=0$, as obtained in Ref.~\cite{Pistolesi-1994} (full line) and by the approximate expression 
 $( 2 \sqrt{2} Q_{c}^{\mathrm{L}} )^{-1}$ (dashed line), with the coupling dependence of $\xi_{\mathrm{K}}^{\sigma}/\pi$ for $\sigma = 20 k_{F}$ (diamonds) taken from Fig.~\ref{Figure-12}.
Although the two quantities $\xi_{\mathrm{pair}}$ and $\xi_{\mathrm{K}}^{\sigma}/\pi$ have been obtained through quite different working procedures, the resemblance here between their coupling dependence appears to be rather remarkable.
This resemblance persists also at finite temperatures up to $T_{c}$, as shown by the results presented in Fig.~\ref{Figure-12}.

To summarize, the above results are all consistent with one's physical expectation that the range of the kernel $K(R)$ of the non-local gap equation (\ref{non-local-LPDA-equation}) should (at any temperature)
be directly related to the size of the Cooper pairs, which represents the fundamental length scale of the BCS pairing theory for fermionic superfluidity \cite{deGennes-1966}.

\vspace{-0.2cm}
\section{Numerical solution of the NLPDA equation for an isolated vortex} 
\label{sec:solving-NLPDA-equation}
\vspace{-0.3cm}

The validity of the (differential) LPDA equation was tested in Ref.~\cite{Simonucci-2014}  for an isolated vortex embedded in an infinite superfluid, for which an accurate solution of the BdG equations is available to compare with across the whole BCS-BEC crossover for all $T < T_{c}$ \cite{Simonucci-2013}.
In Ref.~\cite{Simonucci-2014} deviations between the LPDA and BdG calculations were found in the BCS regime at low temperature, and their origin was attributed to the finiteness of the  spatial range of the kernel of the (integral) NLPDA equation from which the LPDA equation was obtained in Ref.~\cite{Simonucci-2014} at a final step.
No detailed analysis, however, was made in Ref.~\cite{Simonucci-2014} about the consequences of the finiteness of this spatial range, when solving for the gap parameter with a nontrivial spatial profile.

Here, we consider again the case study of an isolated vortex embedded in an infinite superfluid, for which the results of the NLPDA equation (\ref{non-local-LPDA-equation}) can be tested against the results of the BdG equations and also compared with the results of the LPDA equation, over an extended region of the coupling-vs-temperature phase diagram.
To this end, a strategy needs to be implemented to solve numerically the NLPDA equation in an efficient way.

\vspace{0.05cm}
\begin{center}
{\bf A. Vortex solution}
\end{center}
\vspace{-0.2cm}

It was shown in Section~\ref{sec:Kernel-NLPDA} (cf. Figs.~\ref{Figure-2} and \ref{Figure-3}) that the kernel of the NLPDA equation (\ref{non-local-LPDA-equation}) has a rather simple form in $Q$-space. 
This kernel starts at small $Q$ as an inverted parabola with coefficients given by Eqs.~(\ref{non-local-LPDA-equation-uniform}) and (\ref{second-derivative-Q}) and ends up at large $Q$ with the linear behaviour (\ref{large_Q-remaining_part}), and in between has a kink singularity at the critical wave vector $Q_{c}$ (cf. Fig.~\ref{Figure-1}) for $T=0$ and $\mu >0$.
Due to this kink singularity, the Fourier transform of this kernel in $R$-space has an oscillatory behaviour with a slowly decaying tail for large $R$, while the 
large-$Q$ behaviour (\ref{large_Q-remaining_part}) results in a strong singularity at $R=0$.

To avoid those features that can cause problems in the numerical solution of the integral equation, out of the two versions in which this equation can be written
according to the definition (\ref{Kernel-R}), namely,
\begin{eqnarray}
- \frac{m}{4 \pi a_{F}} \, \Delta(\mathbf{r}) & =  & \! \int \! d \mathbf{R} \,\, \Delta(\mathbf{R}) \, K(\mathbf{r}-\mathbf{R}|\mathbf{r}) 
\nonumber \\
& = & \int \! \frac{d\mathbf{Q}}{\pi^{3}} \, e^{2 i \mathbf{Q} \cdot \mathbf{r}} \, \Delta(\mathbf{Q}) \, K(\mathbf{Q}|\mathbf{r})
\label{two-versions-NLPDA-equation}
\end{eqnarray}
\noindent 
it appears convenient to use the second version in $\mathbf{Q}$-space.
[According to the arguments of Section~\ref{sec:Kernel-NLPDA}, it is understood that the modified kernel $K^{\sigma}$ of Eq.~(\ref{kernel-new-definition}) enters Eq.~(\ref{two-versions-NLPDA-equation}), although this step is not strictly necessary when solving the gap equation where $\Delta(\mathbf{Q})$ limits the relevant range of $K(\mathbf{Q}|\mathbf{r})$ to values of $|\mathbf{Q}|$ much smaller than 
$\sigma$.
In addition, the suffix $\mathbf{A}$ has been dropped from Eq.~(\ref{two-versions-NLPDA-equation}) since we consider here the case with $\mathbf{A}=0$.]

To achieve self-consistency of the solution, the choice of the $\mathbf{Q}$-version of Eq.~(\ref{two-versions-NLPDA-equation}) requires us to transform the profile of the gap parameter $\Delta$ back and forth from $\mathbf{R}$- to $\mathbf{Q}$-space as many times as needed.
To this end, an efficient method is required not to loose numerical accuracy in the course of the repeated transformations.
A method to fulfil this purpose is described in detail in Appendix~\ref{sec:appendix-B}.

\begin{figure*}[t]
\begin{center}
\includegraphics[width=14.0cm,angle=0]{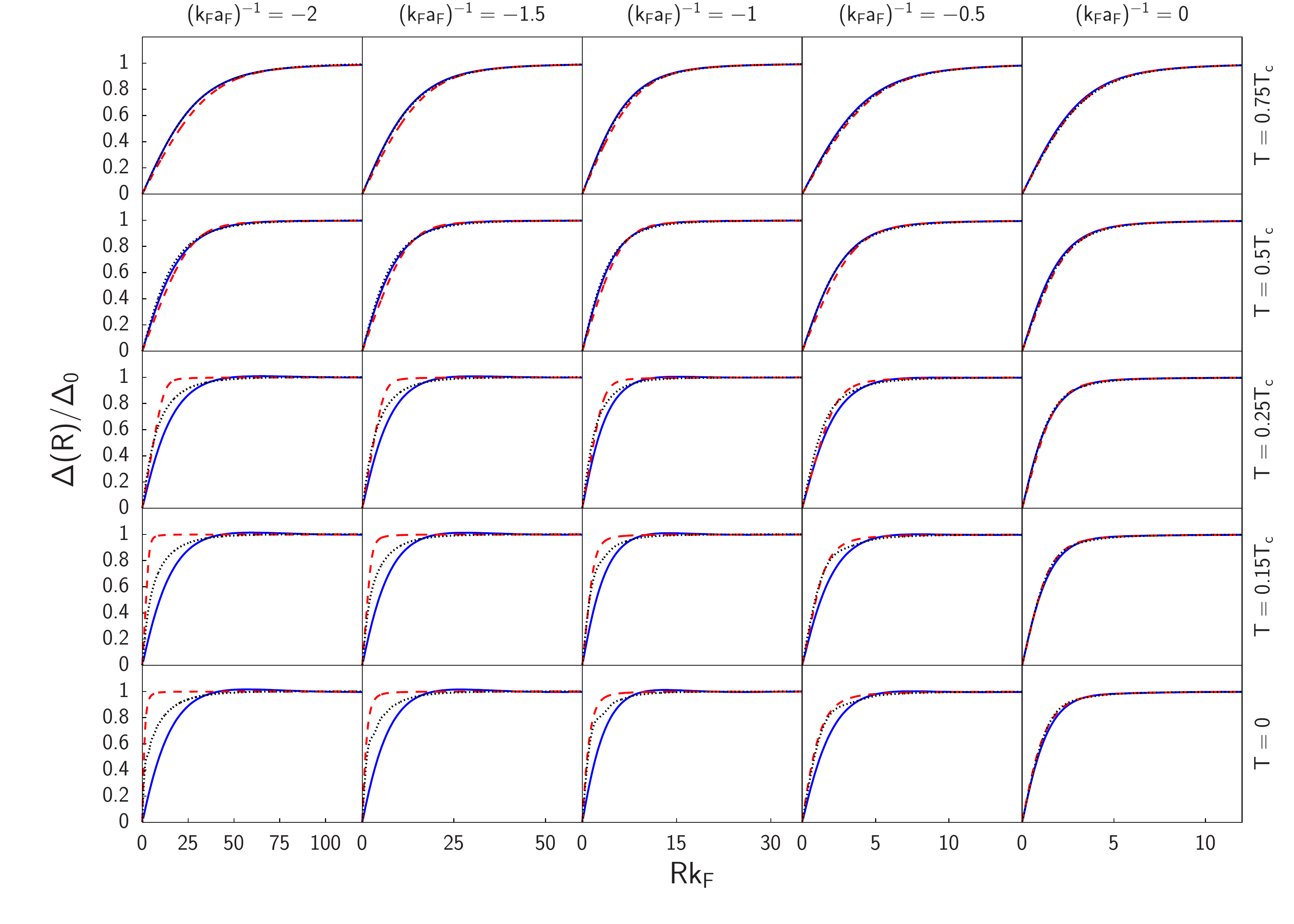}
\caption{(Color online) The radial profiles of the gap parameter $\Delta(R)$ for an isolated vortex (in units of the bulk value $\Delta_{0}$) are shown vs $Rk_{F}$, for the coupling values
                                     $(k_{F}a_{F})^{-1} = (-2.0,-1.5,-1.0,-0.5,0.0)$ and temperatures $T = (0.0,0.15,0.25,0.50,0.75)T_{c}$. 
                                     In each case, the values of $T_{c}$ correspond to the given coupling, while the values of $\Delta_{0}$ correspond to the given coupling and temperature. 
                                     In each panel, three different calculations are reported: NLPDA (full lines); LPDA (broken lines); BdG (dotted lines).}
\label{Figure-14}
\end{center}
\end{figure*}

This method is especially suited when the symmetry of the problem reduces the $\mathbf{R}$- and $\mathbf{Q}$-integration to one dimension.
This is the case of an isolated vortex with cylindrical symmetry embedded in an infinite superfluid, for which the gap parameter takes the form:
\begin{equation}
\Delta(\mathbf{R}) = \Delta(\rho,\varphi,R_{z}) = \Delta(\rho) \, e^{i \varphi}
\label{cylindrical-symmetry}
\end{equation}
\noindent
where $\rho = \sqrt{R_{x}^{2}+R_{y}^{2}}$.
Its Fourier transform then reads:
\begin{equation}
\Delta(\mathbf{Q}) =  \! \int \! d \mathbf{R} \,\, e^{-2i\mathbf{Q}\cdot\mathbf{R}} \Delta(\mathbf{R}) = \pi \delta(Q_{z}) \Delta(Q) \, e^{i \varphi_{Q}}
\label{gap-Fourier-transform}
\end{equation}
\noindent
where $Q=\sqrt{Q_{x}^{2}+Q_{y}^{2}}$ and $\varphi_{Q}$ are the radial and azimuthal coordinates in the $(Q_{x},Q_{y})$ plane.

The results of the NLPDA calculation are reported in Fig.~\ref{Figure-14} for several couplings and temperatures throughout the BCS-BEC crossover, where they are also compared with the corresponding results of the LPDA and BdG approaches that were obtained in Refs.~\cite{Simonucci-2014} and \cite{Simonucci-2013}, respectively.
From these plots one sees that the results of the NLPDA and LPDA calculations coincides with each other and also with those of the BdG calculation over most part of the coupling-vs-temperature phase diagram of the BCS-BEC crossover, with the exception of the BCS side of unitarity at low temperature where deviations occur among the three calculations.
Apart from these deviations in a restricted region of the phase diagram, the results shown in Fig.~\ref{Figure-14} are computationally remarkable, because a considerable gain in memory storage (obtained by the NLPDA and LPDA calculations with respect to the BdG calculation) is accompanied by a large reduction of computational time (by a factor of about $10^{2}$ from the BdG to the NLPDA calculations, and by an additional factor of $10^{3}$ from the NLPDA to the LPDA calculations).

\begin{figure}[t]
\begin{center}
\includegraphics[width=7.0cm,angle=0]{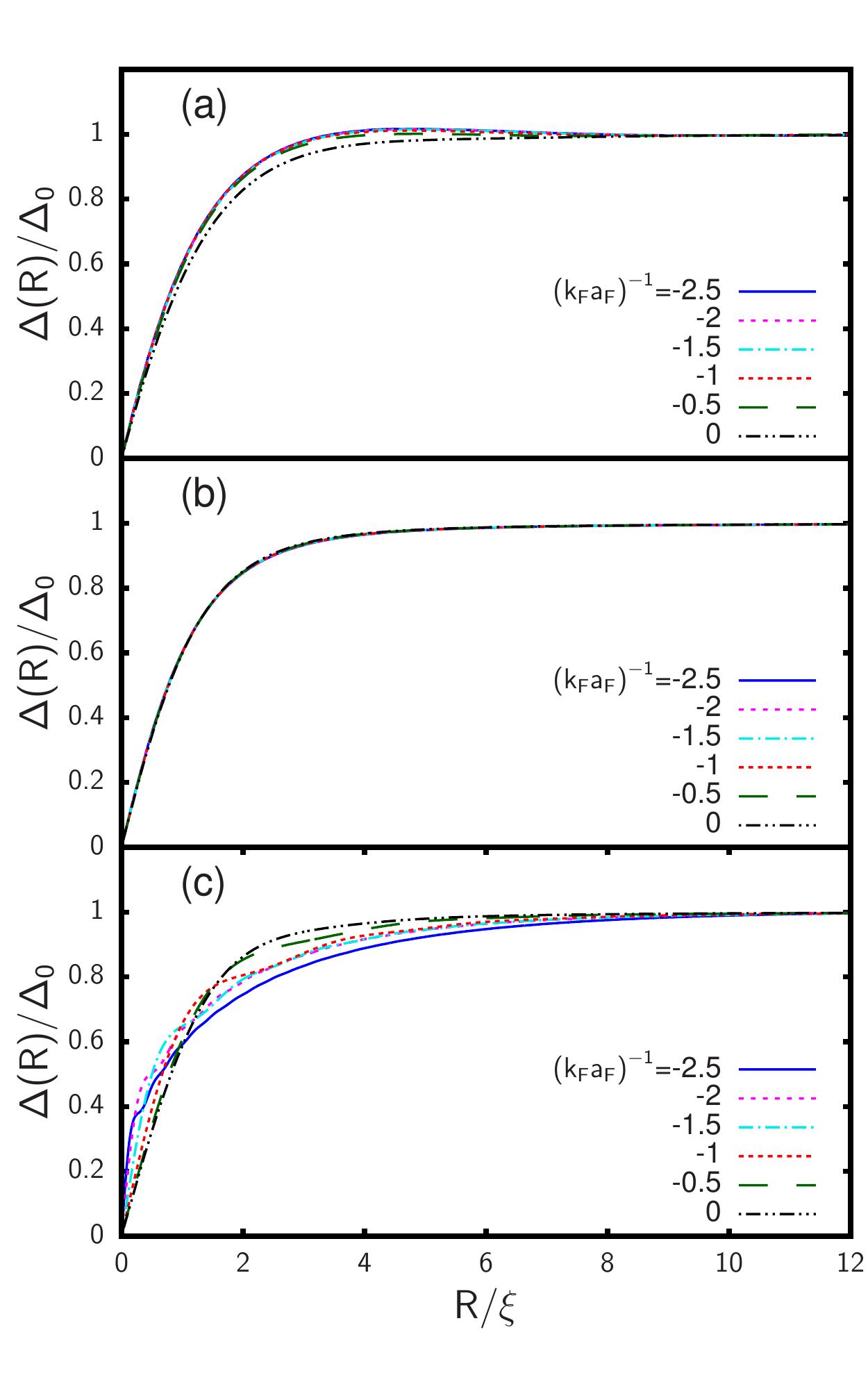}
\caption{(Color online) The radial profiles of the gap parameter $\Delta(R)$ for an isolated vortex (in units of the bulk value $\Delta_{0}$) at $T=0$ are shown for different couplings,
                                     separately for the (a) NLPDA, (b) LPDA, and (c) BdG calculations.
                                     Here, $R$ is in units of a suitably determined coupling-dependent length $\xi$, in such a way that all curves in a given panel fall as close as possible into a single profile.}
\label{Figure-15}
\end{center}
\end{figure}

The deviations occurring in Fig.~\ref{Figure-14} on the BCS side of unitarity at low temperature among the three calculations are interesting and deserve further inquiring.
We note that the NLPDA and LPDA calculations depart from the BdG calculation in a different way.
The NLPDA calculation yields an apparently wider vortex structure than the BdG calculation, while the LPDA calculation yields a narrower vortex structure than the BdG calculation.
To extract from these results information of physical relevance, we have grouped separately the NLPDA, LPDA, and BdG calculations at $T=0$ for different couplings in three separate plots.
In the three panels of Fig.~\ref{Figure-15}, the radial profiles of the gap parameter for various couplings are redrawn by rescaling the spatial coordinate $R$ with respect to a suitably determined length scale $\xi$, in such a way that all profiles fall as close as possible into the shape of a single profile.
This rescaling is seen to work properly for the NLPDA (Fig.~\ref{Figure-15}(a)) and LPDA (Fig.~\ref{Figure-15}(b)) calculations (apart from minor deviations for the NLPDA calculation), thereby implying that \emph{a single kind of length scale} is separately associated with each of these calculations.
On the other hand, it is apparently not possible to obtain a similar result when the rescaling is applied to the BdG radial profiles of the gap parameter for different couplings (Fig.~\ref{Figure-15}(c)).
In this case, the rescaled profiles cross each other at $R \sim k_{F}^{-1}$ with a fan-like shape, showing the presence of \emph{two length scales} which characterize the vortex at short ($\lesssim k_{F}^{-1}$) and large ($\gg k_{F}^{-1}$) distances from its center, respectively.

\begin{figure}[t]
\begin{center}
\includegraphics[width=8.2cm,angle=0]{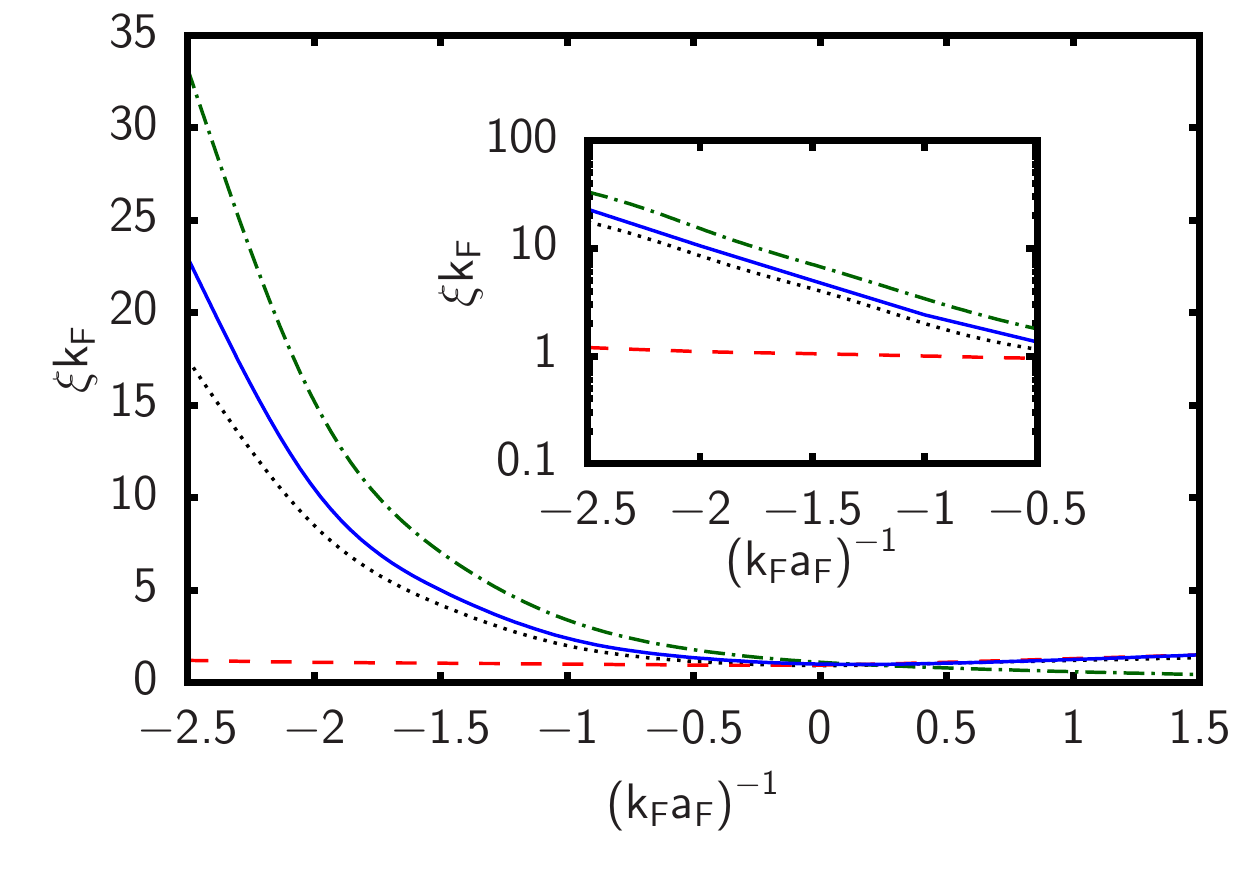}
\caption{(Color online) The length scale $\xi$ (in units of $k_{F}^{-1}$), corresponding to the NLPDA (full line), LPDA (broken line), and BdG (dotted line) calculations reported  
                                     in panels (a), (b), and (c) of Fig.~\ref{Figure-15}, respectively, is shown vs the coupling parameter $(k_{F}a_{F})^{-1}$.
                                     The range of the kernel of the NLPDA equation, given by the Cooper pair size $\xi_{\mathrm{pair}}$, is also shown for comparison (dashed-dotted line).
                                     All results are at $T=0$. In addition, the inset shows the same quantities reported on a semi-log scale on the BCS side of unitarity.}
\label{Figure-16}
\end{center}
\end{figure}

When the above rescaling procedure is working properly, like for the NLPDA and LPDA calculations, the absolute values of the length scale $\xi$ (in units of $k_{F}^{-1}$) used to obtain the plots of  Fig.~\ref{Figure-15} can be determined by making a fit of the radial profile of the gap parameter at unitarity and then using the rescaled values of $\xi$ obtained above to generate the absolute values of $\xi$ at the remaining couplings.
The results are shown in Fig.~\ref{Figure-16} over an extended coupling range about unitarity.
One sees that the values of $\xi$ obtained at $T=0$ on the BCS side of unitarity by the NLPDA calculation about coincide with the range $\xi_{\mathrm{pair}}$ of the kernel of this equation reported in  Fig.~\ref{Figure-13}.
[In this comparison, one should consider that the range of the kernel was calculated in terms of the homogeneous value $\Delta_{0}$, while the NLPDA calculation takes into account the whole profile 
$\Delta(R)$.]
Past unitarity on the BEC side, $\xi$ obtained by the NLPDA calculation begins to increase while the range $\xi_{\mathrm{pair}}$ of the kernel continues to decrease.
The behaviour of $\xi$ obtained by the NLPDA calculation at $T=0$ is then that expected for the coupling dependence of the healing length associated with \emph{inter}-pair correlations \cite{Pistolesi-1996}, which differs from the coupling dependence of the Cooper pair size $\xi_{\mathrm{pair}}$ associated instead with \emph{intra}-pair correlations \cite{Pistolesi-1994}.
On the other hand, $\xi$ obtained by the LPDA calculation equals $k_{F}^{-1}$ on the whole BCS side up to unitarity, past which it catches on with the results of the NLPDA equation.
It was shown analytically in Ref.~\cite{Simonucci-2014} that the length scale $k_{F}^{-1}$ results from the LPDA equation in the BCS limit when the gap parameter vanishes, like at the center of the vortex.
The problem with the LPDA equation on the BCS side of unitarity at $T=0$ is that the length scale $k_{F}^{-1}$ appears not only near the center of the vortex, but is associated with its whole profile.

Figure~\ref{Figure-16} shows, in addition, the values of $\xi$ obtained by fitting the profiles of the gap parameter obtained by the BdG calculation of Ref.~\cite{Simonucci-2013} at $T=0$ for different couplings.
In this case, to associate a meaningful value of $\xi$ with the BdG calculation, one has to consider also the asymptotic $R^{-2}$ behaviour of the vortex at large distances from its center, where the short length scale $k_{F}^{-1}$ that characterizes the center of the vortex has exhausted its effects.
It turns out that the overall coupling dependence of $\xi$ obtained by the BdG calculation is quite similar to that of the NLPDA calculation, as shown more clearly in the inset of Fig.~\ref{Figure-16} \cite{footnote-2}.
On the BCS side of unitarity at low temperature, the NLPDA calculation thus represents a definite improvement with respect to the LPDA calculation.

\begin{figure}[t]
\begin{center}
\includegraphics[width=6.8cm,angle=0]{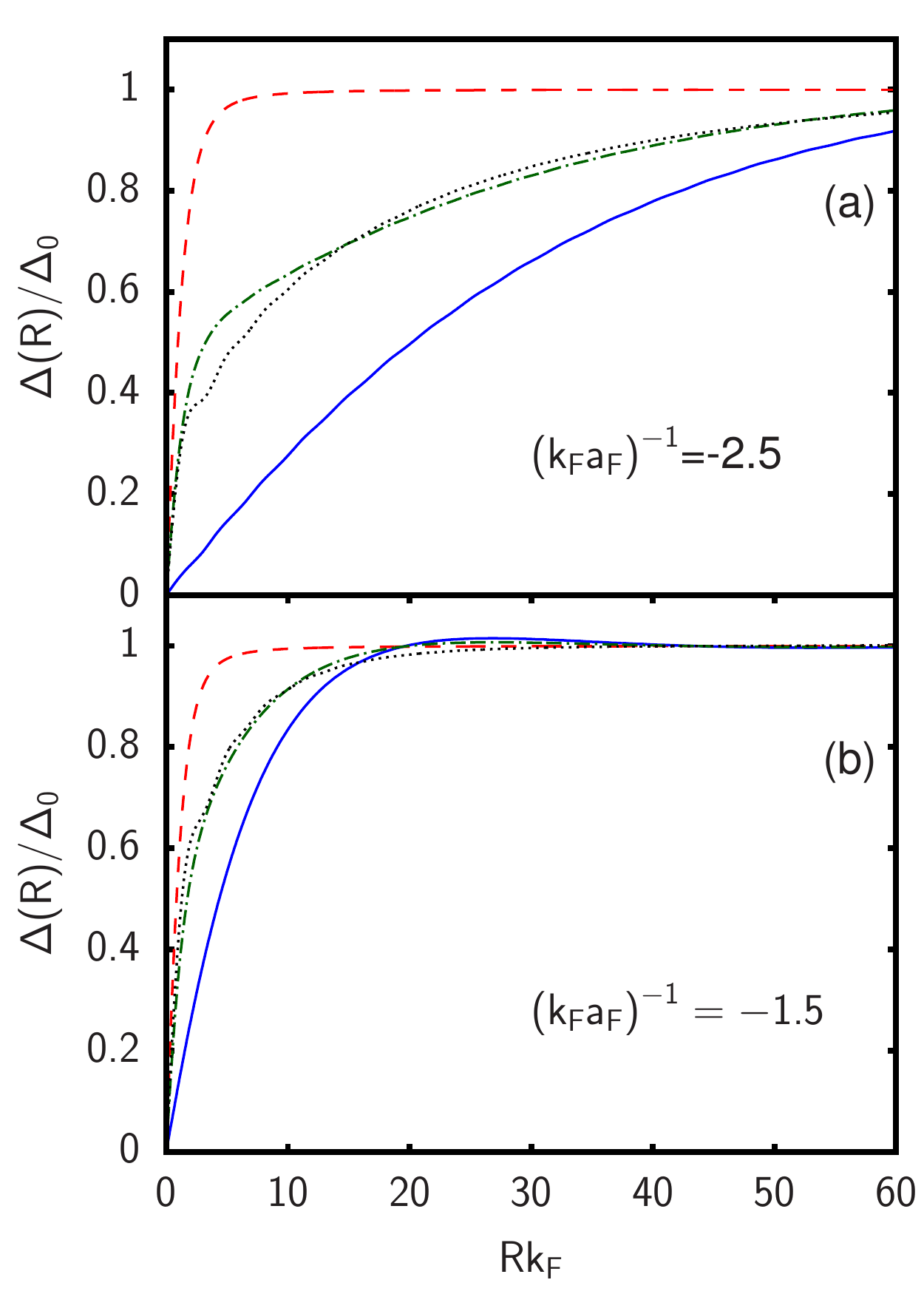}
\caption{(Color online) The radial profiles of the gap parameter $\Delta(R)$ (in units of the bulk value $\Delta_{0}$), obtained at $T=0$ for an isolated vortex by the NLPDA (full line), 
                                     LPDA (dashed line), and BdG (dotted line) calculations for the couplings (a) $(k_{F} a_{F})^{-1}=-2.5$ and (b) $(k_{F} a_{F})^{-1}=-1.5$, are compared with 
                                     the average of the NLPDA and LPDA calculations (dashed-dotted line).}
\label{Figure-17}
\end{center}
\end{figure}

That in this portion of the phase diagram the LPDA and NLPDA calculations reproduce the BdG behavior of the vortex at short and large distances from its center, respectively, can be evidenced in an empirical fashion by averaging the vortex profiles obtained by the two calculations.
This is shown in Fig.~\ref{Figure-17} for the couplings $(k_{F} a_{F})^{-1}=(-2.5,-1.5)$, where good agreement is obtained in this way essentially for all $R$ with the profiles of the BdG calculation.
The plot of Fig.~\ref{Figure-17}(b) also evidences that at $R \sim \xi_{\mathrm{pair}}$ the NLPDA calculation produces a small overshooting of $\Delta(R)$ 
(of the order of a few percents) over the bulk value $\Delta_{0}$.
This small overshooting (which is also apparent in Fig.~\ref{Figure-15}(a)) is inherited from the oscillatory behavior of the kernel $K(\mathbf{R})$ of the NLPDA equation (as shown, for instance, in  Fig.~\ref{Figure-11}).
 
\vspace{0.05cm}
\begin{center}
{\bf B. Granularity scale of the NLPDA equation and validity of the LPDA equation obtained as an approximation to the NLPDA equation}
\end{center}

In Section~\ref{sec:Kernel-NLPDA}, the range of the kernel $K(\mathbf{R}|\mathbf{r})$ of the $\mathbf{R}$-version of the (integral) NLPDA equation (\ref{two-versions-NLPDA-equation}) was shown to be associated with the Cooper pair size $\xi_{\mathrm{pair}}$.
Accordingly, the spatial extent of inhomogeneities (over and above a uniform background) occurring in a given solution $\Delta(\mathbf{r})$ of the NLPDA equation cannot be smaller than $\xi_{\mathrm{pair}}$ itself, which thus turns out to be \emph{the length scale of the ``granularity'' over which the coarse-graining procedure of Ref.\cite{Simonucci-2014} is effective}.
As a consequence, finer details occurring over the smaller fermionic length scale $k_{F}^{-1}$ (which is characteristic of the normal phase but also shows up in the superfluid phase at weak-coupling) are washed out by the coarse-graining procedure through which the NLPDA equation is obtained starting from the BdG equations.
This conclusion was also borne out by the numerical calculations presented in subsection~\ref{sec:solving-NLPDA-equation}-A.

The above considerations can be transferred to the (differential) LPDA equation, that was obtained in Ref.\cite{Simonucci-2014} from the NLPDA equation by the further approximation of expanding its kernel $K(\mathbf{Q}|\mathbf{r})$ about $\mathbf{Q}=0$ up to quadratic order in $\mathbf{Q}$.
To be physically meaningful, a given solution $\Delta(\mathbf{r})$ of the LPDA equation should then contain inhomogeneities (over and above a uniform background) which also have a spatial extent not smaller than the coarse-graining length $\xi_{\mathrm{pair}}$.
As a consequence, whenever this condition is violated in numerical calculations based on the LPDA equation and a smaller spatial extent is instead obtained, the use of the LPDA equation is not justified on physical grounds. 
This is what happens in the weak-coupling (BCS) regime at zero temperature, as already discussed in Ref.\cite{Simonucci-2014} and explicitly considered also in subsection~\ref{sec:solving-NLPDA-equation}-A.

Alternatively, the $\mathbf{Q}$-version of the (integral) NLPDA equation represented by the right-hand side of Eq.~(\ref{two-versions-NLPDA-equation}) can be analyzed to establish the validity of the (differential) LPDA equation. 
At zero temperature, an expansion of the kernel $K(\mathbf{Q}|\mathbf{r})$ about $\mathbf{Q}=0$ up to quadratic order in $\mathbf{Q}$ is expected to hold, \emph{provided} the spread $\delta_{Q}$ of wave vectors about $\mathbf{Q}=0$, over which the solution $\Delta(\mathbf{Q})$ of the NLPDA equation is approximately localized, does not reach the kink singularity at $Q_{c}$ of the kernel $K(\mathbf{Q}|\mathbf{r})$ for  $\mu > 0$.
To determine in practice the values of $\delta_{Q}$, however, care must be exerted in filtering out the numerical oscillations present in the profiles of both $\Delta(R)$ and $\Delta(Q)$ (which originate from different reasons).
With reference to the vortex solution obtained at zero temperature in subsection \ref{sec:solving-NLPDA-equation}-A, for given coupling we have then adopted the following procedure:
(i) The numerical noise that occurs in the profile of $\Delta(R)$ is first smoothed out, by fitting $\Delta(R)$ with the expression $\Delta(R) = P(R) / \sqrt{1 + Q(R)}$ where $P(R)=\sum_{i=0}^{2} p_{i} R^{2i+1}$ and $Q(R)=\sum_{j=1}^{5} q_{j} R^{2j}$ are polynomials with free parameters $\{p_{i},q_{j}\}$;
(ii) The smooth profile of $\Delta(R)$ obtained in this way is multiplied by $\exp{\{-R^{2}/\Gamma^{2}\}}$, in order to focus directly on the overall behaviour of the envelope of $\Delta(Q)$ and avoid dealing with the oscillatory behaviour of $\Delta(Q)$ which would otherwise persists at large $Q$ (owing to the flatness of $\Delta(R)$ when approaching the bulk region away from the vortex centre);
(iii) The Fourier transform of this product is calculated for increasing values of $\Gamma$ until its shape gets stabilized in the ``outer'' region where $Q \gtrsim 5 \, \Gamma^{-1}$
(typically, the value $\Gamma = 5 k_{F}^{-1}$ proves sufficient to the purpose); 
(iv) In the outer region $5 \, \Gamma^{-1} \lesssim Q \lesssim  8 k_{F}$ the resulting shape of this Fourier transform is fitted by the expression $A \exp{\{-Q/\delta_{Q}\}}/Q$, 
to extract the desired value of $\delta_{Q}$.

\begin{figure}[h]
\begin{center}
\includegraphics[width=8.8cm,angle=0]{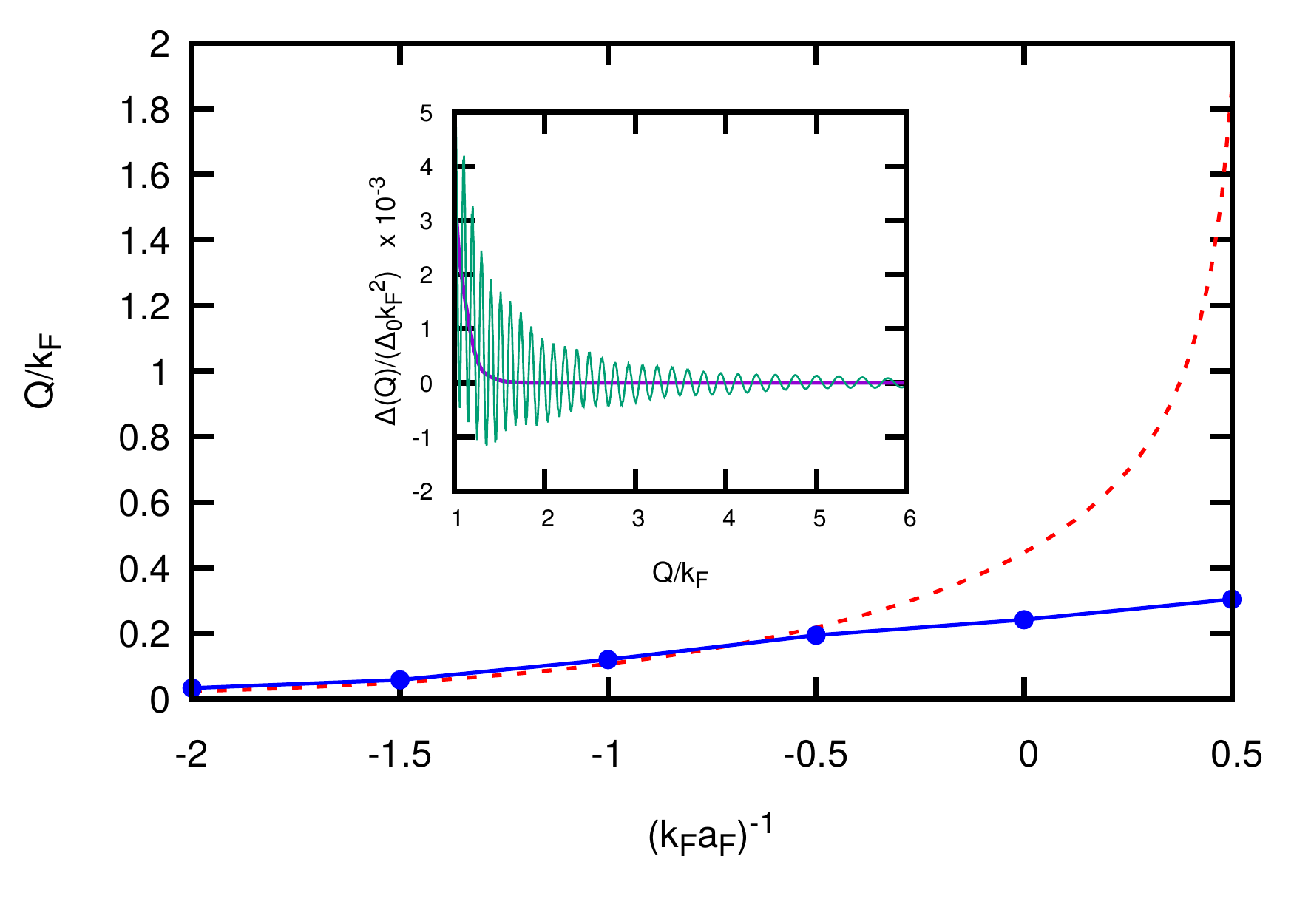}
\caption{(Color online) The spread $\delta_{Q}$ (in units of $k_{F}$) of the vortex solution $\Delta(Q)$ of the gap parameter is shown at $T=0$ as a function of coupling $(k_{F} a_{F})^{-1}$ 
                                    (dots and full line) and compared with the critical wave vector $Q_{c}$ from Fig.~\ref{Figure-1} (dashed line).
                                    The inset shows the profile of $\Delta(Q)$ for $(k_{F} a_{F})^{-1}=-1.0$ (light full line), together with the overall shape of its envelope (heavy full line) used to extract the value 
                                    of $\delta_{Q}$.}
\label{Figure-18}
\end{center}
\end{figure}

The results for the spread $\delta_{Q}$ obtained in this way are shown Fig.~\ref{Figure-18} for several couplings 
(mostly on the BCS side of unitarity), where they are compared with the wave vector $Q_{c}$ from Fig.~\ref{Figure-1}.
Note that, when approaching the weak-coupling (BCS) limit where $Q_{c} \ll k_{F}$, $\delta_{Q}$ does reach (but never exceeds) the value of $Q_{c}$ where a kink singularity occurs in the kernel $K(\mathbf{Q}|\mathbf{r})$ \cite{footnote-3}.
In addition, the inset of Fig.~\ref{Figure-18} shows a typical example of the oscillatory behaviour that affects
$\Delta(Q)$ if step (ii) above would not be implemented, together with the overall shape of the envelope of $\Delta(Q)$ that results once step (ii) is instead adopted.
From this plot one concludes that the validity of the quadratic expansion (from which the LPDA equation is derived from the NLPDA equation) shrinks to progressively smaller values of $|\mathbf{Q}|$ upon approaching the weak-coupling (BCS) limit where $Q_{c} \ll k_{F}$.

When translated back to $\mathbf{R}$-space, the above argument implies that a given solution $\Delta(\mathbf{r})$ of the LPDA equation can be regarded to be physically meaningful, provided that it is spread over a length scale $\ell$ not smaller than $Q_{c}^{-1}$.
Here, the length scale $\ell$ is of the order of the (temperature-dependent) healing length $\xi$ for inter-pair correlations, while $Q_{c}^{-1}$ about coincides with the Cooper pair size $\xi_{\mathrm{pair}}$ associated with intra-pair correlations \cite{Pistolesi-1994,Pistolesi-1996,Palestini-2014}.
From Fig.~\ref{Figure-13} this again implies that, for the LPDA equation to hold, $\ell$ (and thus $\xi$) should not be appreciably smaller than the spatial range $\xi_{\mathrm{pair}}$ of the kernel $K(\mathbf{R}|\mathbf{r})$ in $\mathbf{R}$-space.
This is consistent with the results obtained in subsection~\ref{sec:solving-NLPDA-equation}-A by analyzing the vortex solution of the NLPDA equation.
At finite temperature, on the other hand, the situation is much improved, since the kink singularity of the kernel $K(\mathbf{Q}|\mathbf{r})$ is progressively smoothed out and $\ell$ (and thus $\xi$) readily becomes larger than $\xi_{\mathrm{pair}}$.
At the same time, the fermionic length scale $k_{F}^{-1}$ which is characteristic of the normal phase looses progressively its importance, to the extent that the Fermi surface gets also blurred by temperature effects.

\section{Concluding remarks and perspectives}
\label{sec:conclusions}
\vspace{-0.3cm}

The results obtained in this paper complement and extend the results previously obtained in Ref.~\cite{Simonucci-2014}, where a double coarse-graining procedure was introduced on the BdG equations
to end up eventually with the local (differential) LPDA equation for the gap parameter.
This equation is similar in spirit to the GL and GP equations, but expands their range of validity over a much wider portion of the coupling-vs-temperature phase diagram in the superfluid phase.
In Ref.~\cite{Simonucci-2014} the question of the spatial extent of the ``granularity'' associated with the coarse-graining procedure was, however, left open, since most attention was concentrated there in studying the outcomes of LPDA equation itself.

In the present paper we have considered this question in detail, by studying the properties of the non-local (integral) NLPDA equation that was also reported in Ref.~\cite{Simonucci-2014} at an intermediate step of the derivation of the local (differential) LPDA equation from the BdG equations.
We have found that the spatial extent of the granularity of the coarse-graining procedure is determined by the range of the kernel of the NLPDA integral equation, and that this range depends markedly on coupling but only weakly on temperature, just in the way as the Cooper pair size does.
Application of the NLPDA equation, to determine the profile of an isolated vortex embedded in an infinite superfluid, has further clarified the nature of the length scales that are associated with a vortex by the three independent BdG, LPDA, and NLPDA approaches for different couplings and temperatures.

Accordingly, we have found that the double coarse-graining procedure of Ref.~\cite{Simonucci-2014} leads to a granularity scale given by the Cooper pair size, in such a way that the internal wave function of the pair becomes irrelevant.
This procedure then effectively averages out the fast oscillations occurring on the scale of $k_{F}^{-1}$, which would anyway provide redundant information when interested in superconductivity.
The coarse-graining procedure of Ref.~\cite{Simonucci-2014} is thus similar in spirit to the Eilenberger quasi-classical approach, which averages out the ``fast'' oscillations in the relative coordinate and retains
the ``slow'' oscillations associated with the center-of-mass coordinate of a pair \cite{Eilenberger-1968}.
However, while the Eilenberger is limited to the weak-coupling (BCS) regime where the underlying Fermi surface plays a dominant role, with the present NLPDA approach it is possible to span the whole
BCS-BEC crossover.

The non-local (integral) NLPDA equation may give ready access to problems that are difficult to deal with using the (local) LPDA equation.
In particular, where the non-local (integral) NLPDA equation is expected to have its most exclusive applications is in the context of the proximity effect, for which the finite size of Cooper pairs plays a key role and cannot be dealt with by using a local (differential) equation \cite{deGennes-1966}.
Specifically, one can consider a surface problem arising at the interface between two superconductors with different couplings (and thus with different critical temperatures), 
such that the paired state in the superconductor at the left ($L$) of the interface kept a temperature $T$ below its critical temperature $T_{c}^{L}$ leaks to the superconductor 
at the right ($R$) of the interface for which $T$ is larger than its critical temperature $T_{c}^{R}$.
This kind of problems was already studied theoretically in Ref.~\cite{Kogan-1982} although only in the weak-coupling (BCS) limit, and can now be carried over to the whole BCS-BEC crossover in terms of the non-local (integral) NLPDA equation.
This study may also help stimulating a revival of the experiments that adopt a similar geometry and physical arrangement, in line with the original experimental work of Ref.~\cite{Polturak-1991} aimed at determining the temperature dependence of the coherence length in the normal phase.


\begin{center}
\begin{small}
{\bf ACKNOWLEDGMENTS}
\end{small}
\end{center}
\vspace{-0.2cm}

We are indebted to P. Pieri for having determined the expression (\ref{large_Q-remaining_part}) for the large-$Q$ behavior of the kernel $K(Q)$ as well as for a critical reading of the manuscript.

\appendix                                                                                                                                                                                                                                                                                                                                                                                                        
\section{ORIGIN OF THE SPATIAL OSCILLATIONS AND THE LONG-RANGE TAIL OF THE KERNEL $\mathbf{K^{\sigma}(R)}$: A MODEL STUDY}
\label{sec:appendix-A}
\vspace{-0.3cm}

The origin of the spatial oscillations, that affect the kernel $K^{\sigma}(R)$ of the non-local gap equation (\ref{non-local-LPDA-equation}) at zero temperature and up to 
$(k_{F} a_{F})^{-1} \lesssim 0.50$ (as shown in Fig.~\ref{Figure-5}), can be identified by considering the following \emph{model function}:
\begin{equation}
K_{\mathrm{model}}(Q) = \left\{ \begin{array}{cc}       0              &   \,\,\, (Q < Q_{0})    \\
                                                                   - \alpha (Q - Q_{0})   &   \,\,\, (Q \ge Q_{0})       \end{array}  \right.      \, .
\label{model-function-Q-space}
\end{equation}
\noindent
This function has a kink at $Q_{0}$ of the same type that the original kernel $K(Q)$ has at the critical wave vector $Q_{c}$ shown in Fig.~\ref{Figure-1}.
To calculate the Fourier transform in $R$-space of the function (\ref{model-function-Q-space}), it is again necessary to multiply it by $e^{-Q^{2}/ \sigma^{2}}$ as we did in 
Eqs.~(\ref{kernel-new-definition}) and (\ref{asymptotic-kernel-new-definition}), and consider the limit $\sigma \rightarrow \infty$  only at the end of the calculation.
 
For given values of $Q_{0}$ and $\sigma$, we write for the Fourier transform in $R$-space of the function (\ref{model-function-Q-space}) (once multiplied by $e^{-Q^{2}/ \sigma^{2}}$):
\begin{equation}
K_{\mathrm{model}}^{\sigma}(R) \, = \, - \, \frac{2 \alpha}{\pi^{2} R} \left\{ J_{2}(R;Q_{0},\sigma) - Q_{0} \, J_{1}(R;Q_{0},\sigma) \right\}
\label{FT-K-model-1}
\end{equation}
\noindent
with the notation
\begin{eqnarray}
J_{1}(R;Q_{0},\sigma) & = & - \sigma^{2} \, \frac{d}{d y} \int_{Q_{0}/\sigma}^{\infty} \!\! dx \, \cos(x y) \, e^{-x^{2}}
\label{integral-J1-definition} \\
J_{2}(R;Q_{0},\sigma) & = & - \sigma^{3} \, \frac{d^{2}}{d y^{2}} \int_{Q_{0}/\sigma}^{\infty} \!\! dx \, \sin(x y) \, e^{-x^{2}}
\label{integral-J2-definition}
\end{eqnarray}
\noindent
where $y = 2 \sigma R$.
The integrals (\ref{integral-J1-definition}) and (\ref{integral-J2-definition}) can then be calculated in a closed form for all values of $y$ (and thus of $R$), in terms of the error function erf($z$) 
of complex argument $z$ \cite{Abramowitz-Stegun}.
Specifically, we can make use of the indefinite integrals reported in Ref.\cite{Ng-Geller-1969}
\begin{footnotesize}
\begin{eqnarray}
\mathrm{erf}(x + i y/2) \, + \mathrm{erf}(x - i y/2) & = & \frac{4 \, e^{y^{2}/4}}{\sqrt{\pi}} \int \! dx \cos(x y) \, e^{-x^{2}}
\label{integrals-error-function-plus} \\
\mathrm{erf}(x + i y/2) \, - \mathrm{erf}(x - i y/2) & = & \frac{4 \, e^{y^{2}/4}}{i \, \sqrt{\pi}} \int \! dx \sin(x y) \, e^{-x^{2}} 
\label{integrals-error-function-minus}
\end{eqnarray}
\end{footnotesize}

\noindent
and reduce the integrals (\ref{integral-J1-definition}) and (\ref{integral-J2-definition}) to the following expressions:
\begin{small}
\begin{eqnarray}
& & J_{1}(R;Q_{0},\sigma) = - \frac{\sqrt{\pi} \sigma^{2}}{4} 
\nonumber \\
& \times & \frac{d}{d y} \!\!  \left\{ \! e^{-\frac{y^{2}}{4}} \!\! \left[ \mathrm{erfc} \! \left( \! \frac{Q_{0}}{\sigma} \! + \! i \frac{y}{2} \right) 
                                                                                                                          \! + \! \mathrm{erfc} \! \left( \! \frac{Q_{0}}{\sigma} \! - \! i \frac{y}{2} \right) \! \right] \! \right\}
\label{integral-J1-solution}  \\
& & J_{2}(R;Q_{0},\sigma) = - \frac{i \, \sqrt{\pi} \sigma^{3}}{4} 
\nonumber \\
& \times & \frac{d^{2}}{d y^{2}} \!\! \left\{ \! e^{-\frac{y^{2}}{4}} \!\! \left[ \mathrm{erfc} \! \left( \! \frac{Q_{0}}{\sigma} \! + \! i \frac{y}{2} \right) 
                                                                                                                          \! - \! \mathrm{erfc} \! \left( \! \frac{Q_{0}}{\sigma} \! - \! i \frac{y}{2} \right) \! \right]  \!\right\}
\label{integral-J2-solution} 
\end{eqnarray}
\end{small}

\noindent
where again $y=2 \sigma R$ and $\mathrm{erfc}(z) = 1 - \mathrm{erf}(z)$ is the complementary error function \cite{Abramowitz-Stegun}.
In particular, the expressions (\ref{integral-J1-solution}) and (\ref{integral-J2-solution}) can be readily calculated for $y \gg 1$ and $y \ll 1$, to obtain the large-$R$ and small-$R$ behaviours of the model 
function (\ref{FT-K-model-1}), respectively.

For $z \rightarrow \infty$, the following asymptotic expansion of the error function can be used \cite{Abramowitz-Stegun}:
\begin{equation}
\mathrm{erfc}(z) \simeq \frac{e^{-z^{2}}}{\sqrt{\pi} \, z} \left( 1 - \frac{1}{2 \, z^{2}} + \frac{3}{4 \, z^{4}} + \cdots \right) 
\label{asymptotic-expansion-erfc}
\end{equation}
\noindent
which is valid for $|\mathrm{arg}z| < 3 \pi/4$ (this condition is satisfied in our case since in the expressions (\ref{integral-J1-solution}) and  (\ref{integral-J2-solution}) $Q_{0}$ is positive).
We thus obtain for $R \rightarrow \infty$:
\begin{small}
\begin{equation}
K_{\mathrm{model}}^{\sigma}(R \rightarrow \infty) \, \simeq \, \frac{\alpha}{2 \, \pi^{2}} \left[ \frac{Q_{0}}{R^{3}} \, \sin (2 Q_{0} R) + \frac{1}{R^{4}} \, \cos (2 Q_{0} R) \right]
\label{K-model-large-R}
\end{equation}
\end{small}

\noindent
where on the right-hand side the limit $\sigma \rightarrow \infty$ has be taken.
Provided that $Q_{0} \ne 0$, the expression (\ref{K-model-large-R}) shows an oscillatory behaviour with wave vector $2 Q_{0}$ and an amplitude that decays like $R^{-3}$ for large $R$.
We thus conclude that it is the kink at $Q_{0}$ of the model function (\ref{model-function-Q-space}) to be responsible both of the oscillatory behaviour and the slow decay of its Fourier transform for large $R$.

For $z \rightarrow 0$, we exploit the fact that the error function $\mathrm{erf}(z)$ is an entire function in the complex $z$-plane, so that its Taylor series always converges, and
use the following expression for its derivatives \cite{Abramowitz-Stegun}:
\begin{eqnarray}
\frac{d^{n+1}}{d z^{n+1}} \mathrm{erfc}(z) & = & - \frac{d^{n+1}}{d z^{n+1}} \mathrm{erf}(z) 
\nonumber \\
& = & (-1)^{n+1} \frac{2}{\sqrt{\pi}} \, H_{n}(z) \, e^{- z^{2}}
\label{derivative-erfc}
\end{eqnarray}

\noindent
where $n = 0,1,2,\cdots$ and $H_{n}(z)$ is the Hermite polynomial of index $n$.
To the leading order in the small parameter $Q_{0}/\sigma$, we thus obtain for $R \rightarrow 0$:
\begin{equation}
K_{\mathrm{model}}^{\sigma}(R \rightarrow 0) \, \simeq \, - \frac{2 \,\alpha \, \sigma^{4}}{\pi^{2}} \left( 1 - \frac{Q_{0}}{2 \, \sigma} \right)
\label{K-model-small-R}
\end{equation}
\noindent
which diverges in the limit $\sigma \rightarrow \infty$.
It was verified numerically in Fig.~\ref{Figure-6} of the main text that this is precisely the kind of divergence that occurs in the kernel $K^{\sigma}(R)$
of the non-local gap equation (\ref{non-local-LPDA-equation}) for $R \rightarrow 0$.
    
\vspace{0.3cm}                                                                                                                                                                                                                                                                                                                                                                                                         
\section{METHOD FOR THE NUMERICAL SELF-CONSISTENT SOLUTION OF THE NON-LOCAL GAP EQUATION}
\label{sec:appendix-B}
\vspace{-0.3cm}

In this Appendix, a method is set up for the numerical self-consistent solution of the non-local gap equation, in the form of Eq. (\ref{two-versions-NLPDA-equation}).
This method amounts to calculating the Fourier transform of the gap parameter $\Delta$ back and forth from $\mathbf{R}$- to $\mathbf{Q}$-space in an efficient way, thus
enabling one to solve for $\Delta$ in $\mathbf{Q}$-space and transferring the information to $\mathbf{R}$-space.
This efficiency reflects itself in the fact that the double Fourier transform (from $\mathbf{R}$-space to $\mathbf{Q}$-space and then back to $\mathbf{R}$-space) is essentially exact, in the sense
that it does not introduce numerical noise since it is based on an orthogonal transformation (cf. Eq.~(\ref{S-matrix}) below).
The present method, which does not rely on the widely applied Fast Fourier Transform method \cite{Brigham-1988}, is especially useful when the gap parameter has special symmetries
like in the case of the cylindrical vortex considered in subsection~\ref{sec:solving-NLPDA-equation}-A.
In essence, this new method rests on appropriately combining the following \emph{two properties} of mathematical physics.

The \emph{first property} refers to the quantum harmonic oscillator in $D$-dimensions (in units $m=1$, $\omega=1$, and $\hbar=1$).
According to this property, if $\psi(\mathbf{r})$ is eigen-function of the Hamiltonian $(-\nabla_{\mathbf{r}}^{2} + \mathbf{r}^{2})/2$ in real space $\mathbf{r}$ with eigenvalue $\varepsilon$, then its Fourier transform $\tilde{\psi}(\mathbf{k})$ is eigen-function of the corresponding Hamiltonian $(-\nabla_{\mathbf{k}}^{2} + \mathbf{k}^{2})/2$ in wave-vector space $\mathbf{k}$ with the same eigenvalue 
$\varepsilon$.
[For clarity, in this Appendix we identify the Fourier transform $\tilde{f}(\mathbf{k})$ of a function $f(\mathbf{r})$ by adding a tilde over its symbol.]
With the notation (\ref{gap-Fourier-transform}) for the Fourier transform, we thus have that $\tilde{\psi}(\mathbf{k}) = (2 \pi)^{D/2} \gamma \, \psi(2 \mathbf{k})$ where $\gamma$ is a complex factor with unit magnitude 
and $\psi(\mathbf{k})$ has the \emph{same} form of $\psi(\mathbf{r})$ with the variable $\mathbf{k}$ replacing $\mathbf{r}$.

In particular, the symmetry of the harmonic potential $\mathbf{r}^{2}/2$ can be exploited to express $\psi(\mathbf{r})$ as the product $\mathcal{R}_{nl}(r) \mathcal{Y}_{lm}(\hat{r})$ with
$r=|\mathbf{r}|$ and $\hat{r}=\mathbf{r}/|\mathbf{r}|$, where $\mathcal{Y}_{lm}(\hat{r})$ is eigen-function of the angular part of the Laplacian in $D$-dimensions.
[For $D=3$ this corresponds to a spherical harmonic $Y_{lm}(\vartheta,\varphi)$, 
for $D=2$ to a planar harmonic $\Phi_{m}(\varphi)= e^{i m \varphi}/\sqrt{2 \pi}$ (with $\ell \leftrightarrow |m|$), 
and for $D=1$ to even ($l \leftrightarrow 0$) and odd ($l \leftrightarrow 1$) parity.]
Quite generally, in $D$-dimensions the radial part $\mathcal{R}_{nl}(r)$ can be expresses in terms of the generalized Laguerre polynomials $\mathcal{L}^{\alpha}_{n}(u)$ \cite{MOS-1966}, in the form:
\begin{equation}
\mathcal{R}_{nl}(r) = \mathcal{N} \, r^{l} \, e^{- r^{2}/2} \, \mathcal{L}^{\alpha}_{n}(r^{2})
\label{radial-part-harmonic-oscillator}
\end{equation}
\noindent
where $\mathcal{N}$ is a normalization factor, $n=0,1,2,\cdots$, and $\alpha= D/2 + l -1$ has a fixed value for given $D$ and $l$.
In what follows, it will be convenient to generalize the expression (\ref{radial-part-harmonic-oscillator}) by considering a $D$-dimensional harmonic oscillator with $m=2 \lambda^{2}$, $\omega=1$, and $\hbar=1$,
whose eigen-functions have the form:
\begin{equation}
\phi^{(\lambda)}_{nlm}(\mathbf{r}) = \mathcal{N}(\lambda) \, (\sqrt{2} \lambda r)^{l} \, e^{- \lambda^{2} r^{2}} \mathcal{L}^{\alpha}_{n}(2 \lambda^{2} r^{2}) \, \mathcal{Y}_{lm}(\hat{r})
\label{gereralized-radial-part-harmonic-oscillator}
\end{equation}
\noindent
where $\mathcal{N}(\lambda)=\sqrt{2} (2 \lambda^{2})^{D/4}$ and again $\alpha= D/2 + l -1$.
The Fourier transform of the function (\ref{gereralized-radial-part-harmonic-oscillator}) reads:
\begin{small}
\begin{equation}
\tilde{\phi}^{(\lambda)}_{nlm}(\mathbf{k}) \! = \! (-i)^{l+2n} \pi^{D/2} \mathcal{N}\left(\!\frac{1}{\lambda}\!\right) \!\! \left(\!\!\frac{\sqrt{2} k}{\lambda}\!\!\right)^{l} \!\!\! e^{- \frac{k^{2}}{\lambda^{2}}} 
\mathcal{L}^{\alpha}_{n}\left(\!\frac{2 k^{2}}{\lambda^{2}}\!\right) \mathcal{Y}_{lm}(\hat{k})
\label{FT-gereralized-radial-part-harmonic-oscillator}
\end{equation}
\end{small}

\noindent
since in this case $\gamma = (-i)^{l+2n}$. 
The parameter $\lambda$ in Eqs.~(\ref{gereralized-radial-part-harmonic-oscillator}) and (\ref{FT-gereralized-radial-part-harmonic-oscillator}) is meant to provide additional flexibility to the numerical calculations. 

The \emph{second property} refers specifically to the generalized Laguerre polynomials $\mathcal{L}^{\alpha}_{n}(u)$.
For given $\alpha$ and varying $n$, these form a family of orthogonal polynomials with respect to the (positive definite) weight function $\rho(u)=u^{\alpha}e^{-u}$, in the sense that:
\begin{equation}
\int_{0}^{\infty} \! \!du \, \rho(u) \, \mathcal{L}^{\alpha}_{n}(u) \, \mathcal{L}^{\alpha}_{n'}(u) = \delta_{n,n'} \, .
\label{orthogonal-polynomials}
\end{equation}
\noindent
In practice, this integral can be represented by a Gaussian quadrature, of the form:
\begin{equation}
\int_{0}^{\infty} \! \! du \, \rho(u) \, \mathcal{L}^{\alpha}_{n}(u) \, \mathcal{L}^{\alpha}_{n'}(u) = \sum_{j=1}^{N} \mathcal{L}^{\alpha}_{n}(u_{j}) \, \mathcal{L}^{\alpha}_{n'}(u_{j}) \, w_{j} = \delta_{n,n'}
\label{orthogonal-polynomials-Gaussian-quadrature}
\end{equation}
\noindent
which is exact for $(n,n') \le N-1$.
Here, the points $\{u_{j}; j=1,\cdots,N\}$ and the associated (positive definite) weights $\{w_{j}; j=1,\cdots,N\}$ have to be suitably determined.
The expression (\ref{orthogonal-polynomials-Gaussian-quadrature}) can also be interpreted as defining a transformation from the $N$ generalized Laguerre polynomials $\mathcal{L}^{\alpha}_{n}(u)$
(with $n=0,1,\cdots,N-1$) to the $N$ points $u_{j}$ (with $j=1,\cdots,N)$ along the $u$-axis, in terms of the orthogonal ($N \times N$) matrix
\begin{equation}
S_{n j} = \mathcal{L}^{\alpha}_{n}(u_{j}) \, \sqrt{w_{j}}
\label{S-matrix}
\end{equation}
\noindent
with given $\alpha$, such that 
\begin{equation}
\sum_{j=1}^{N} S_{n j} S^{T}_{j n'} = \delta_{n,n'} \,\,\,\,\,\, \mathrm{and} \,\,\,\,\,\, \sum_{n=0}^{N-1} S^{T}_{j n} S_{n j'} = \delta_{j,j'} \,\, .
\label{orthogonality-S-matrix}
\end{equation}
\noindent
An efficient method for generating the set of points $u_{j}$ and the matrix elements (\ref{S-matrix}) will be described below. 

The two above properties (represented by Eqs.~(\ref{gereralized-radial-part-harmonic-oscillator}) and (\ref{FT-gereralized-radial-part-harmonic-oscillator}), and by Eqs.~(\ref{S-matrix}) and (\ref{orthogonality-S-matrix}), respectively) can be combined into a method for solving numerically the non-local gap equation (\ref{two-versions-NLPDA-equation}).
To this end, we consider the projection of the gap parameter of the form $\Delta(\mathbf{R}) = \Delta(R) \, \mathcal{Y}_{lm}(\hat{R})$ onto the set of functions $\phi^{(\lambda)}_{nlm}$,
alternatively in $\mathbf{R}$- and $\mathbf{Q}$-space.
In $\mathbf{R}$-space we obtain:
\begin{small}
\begin{eqnarray}
& & \int \! \! d\mathbf{R} \,\, \phi^{(\lambda)}_{nlm}(\mathbf{R})^{*} \Delta(\mathbf{R})
\label{projection-R-space} \\
& = & \frac{1}{\sqrt{2} (2 \lambda^{2})^{D/4}} \! \int_{0}^{\infty} \! \! dx \, x^{\alpha} e^{-x} \mathcal{L}^{\alpha}_{n}(x) x^{-\frac{l}{2}} e^{x/2} \Delta \! \left( \! \sqrt{\frac{x}{2 \lambda^{2}}}\right)
\nonumber \\
& \simeq & \frac{1}{\sqrt{2} (2 \lambda^{2})^{D/4}} \sum_{j=1}^{N} \mathcal{L}^{\alpha}_{n}(x_{j}) x_{j}^{-\frac{l}{2}} e^{x_{j}/2} \Delta \!\left( \! \sqrt{\frac{x_{j}}{2 \lambda^{2}}}\right) w_{j} 
\nonumber
\end{eqnarray}
\end{small}

\noindent
where we have set $x = 2 \lambda^{2} R^{2}$ and the weights $w_{j}$ are associated with the weight function $x^{\alpha} e^{-x}$.
In $\mathbf{Q}$-space we obtain instead:
\begin{small}
\begin{eqnarray}
& & \int \! \! \frac{d\mathbf{Q}}{\pi^{D}} \,\, \tilde{\phi}^{(\lambda)}_{nlm}(\mathbf{Q})^{*} \tilde{\Delta}(\mathbf{Q})
\label{projection-Q-space} \\
& = & \frac{(i)^{l+2n} (\lambda^{2}/2)^{D/4}}{\sqrt{2 \pi^{D}}} \!\! \int_{0}^{\infty} \! \! dx \, x^{\alpha} e^{-x} \mathcal{L}^{\alpha}_{n}(x) x^{-\frac{l}{2}} e^{x/2} \tilde{\Delta} \! \left( \! \sqrt{\frac{x \lambda^{2}}{2}}\right)
\nonumber \\
& \simeq & \frac{(i)^{l+2n} (\lambda^{2}/2)^{D/4}}{\sqrt{2 \pi^{D}}} \sum_{j=1}^{N} \mathcal{L}^{\alpha}_{n}(x_{j}) x_{j}^{-\frac{l}{2}} e^{x_{j}/2} \tilde{\Delta} \! \left( \! \sqrt{\frac{x_{j} \lambda^{2}}{2}}\right) w_{j} 
\nonumber
\end{eqnarray}
\end{small}

\noindent
since $\tilde{\Delta}(\mathbf{Q}) = \tilde{\Delta}(Q) \mathcal{Y}_{lm}(\hat{Q})$ and where we have now set $x = 2 Q^{2}/\lambda^{2}$.
Note that the \emph{same mesh of $x$ points} $\{x_{j}; j=1,\cdots,N\}$ has been used on the right-hand side of Eqs.~(\ref{projection-R-space}) and (\ref{projection-Q-space}), in such a way that $Q_{j} = \lambda^{2} R_{j}$ for each value of $j$.
By this choice, the meshes of $R$ and $Q$ points are interlinked to each other in an appropriate way.
This represents a key property of the method. 
In addition, the possibility of changing the value of $\lambda$ (besides changing the value of $N$) provides some additional flexibility to the numerical calculations.

The expressions (\ref{projection-R-space}) and (\ref{projection-Q-space}) are equal to each other owing to a property of the Fourier transforms.
By equating their right-hand sides, we then write in a compact form:
\begin{small}
\begin{equation}
\sum_{j=1}^{N} S_{n j} \, y_{j} \Delta \!\left( \! \sqrt{\frac{x_{j}}{2 \lambda^{2}}}\right)
\! = \! \frac{(i)^{l+2n} \lambda^{D}}{\sqrt{\pi^{D}}} \sum_{j=1}^{N} S_{n j} \, y_{j} \tilde{\Delta} \!\left( \! \sqrt{\frac{x_{j} \lambda^{2}}{2}}\right)
\label{comparison-two-expressions}
\end{equation}
\end{small}

\noindent
with the matrix elements $S_{n j}$ given by Eq.~(\ref{S-matrix}) and where we have set
\begin{equation}
y_{j} = e^{x_{j}/2} x_{j}^{-l/2} \sqrt{w_{j}} \, .
\label{definition-y}
\end{equation}

At this point, we can extract alternatively the quantities \begin{small}$\Delta \! \left( \! \sqrt{\frac{x_{j}}{2 \lambda^{2}}}\right)$\end{small} and 
\begin{small}$\tilde{\Delta} \! \left( \! \sqrt{\frac{x_{j} \lambda^{2}}{2}}\right)$\end{small}, by multiplying the expression
(\ref{comparison-two-expressions}) by $S^{T}_{j' n}$ and by $(-1)^{n}S^{T}_{j' n}$, respectively, summing over $n$, and taking into account the orthogonality properties (\ref{orthogonality-S-matrix}).
We obtain eventually:
\begin{small}
\begin{equation}
\Delta \! \left( \! \sqrt{\frac{x_{j'}}{2 \lambda^{2}}} \right) = \frac{i^{l} \lambda^{D}}{\sqrt{\pi^{D}} \, y_{j'}} \sum_{j=1}^{N} \sum_{n=0}^{N-1} (-1)^{n} S^{T}_{j' n} S_{n j} \, y_{j} 
\tilde{\Delta} \!\left( \! \sqrt{\frac{x_{j} \lambda^{2}}{2}}\right)
\label{explicit-expression-Delta-R} 
\end{equation}
\end{small}
\noindent
as well as 
\begin{small}
\begin{equation}
\tilde{\Delta} \! \left( \! \sqrt{\frac{x_{j'} \lambda^{2}}{2}} \right)  = \frac{(-i)^{l} \sqrt{\pi^{D}}}{\lambda^{D} \, y_{j'}} \sum_{j=1}^{N} \sum_{n=0}^{N-1} (-1)^{n} S^{T}_{j' n} S_{n j} \, y_{j} 
\Delta \!\left( \! \sqrt{\frac{x_{j}}{2 \lambda^{2}}}\right).
\label{explicit-expression-Delta-Q} 
\end{equation}
\end{small}

\noindent
Note that the presence of the factor $(-1)^{n}$ (which originates from Eq.~(\ref{FT-gereralized-radial-part-harmonic-oscillator})) on the right-hand side of the expressions (\ref{explicit-expression-Delta-R}) and (\ref{explicit-expression-Delta-Q}) is essential to get a meaningful result when summing over $n$.
Recall further that the results (\ref{explicit-expression-Delta-R}) and (\ref{explicit-expression-Delta-Q}) are approximate to the extent that the right-hand sides of Eq.~(\ref{projection-R-space}) 
and (\ref{projection-Q-space}) are also approximate.

We can eventually make use of the results (\ref{explicit-expression-Delta-R}) and (\ref{explicit-expression-Delta-Q}) to cast the $\mathbf{Q}$-version of the NLPDA equation (\ref{two-versions-NLPDA-equation}) into an \emph{algebraic} form.
Since $\tilde{K}(\mathbf{Q}|\mathbf{r})$ depends only on $|\mathbf{Q}|$, the product $\tilde{\Delta}(\mathbf{Q}) \tilde{K}(\mathbf{Q}|\mathbf{r})$ maintains the same symmetry of $\tilde{\Delta}(\mathbf{Q})$.
[Recall that, according to a convention adopted in this Appendix, a tilde is understood to appear on both $\Delta(\mathbf{Q})$ and $K(\mathbf{Q}|\mathbf{r})$ in Eq.~(\ref{two-versions-NLPDA-equation}).]
By applying successively Eqs.~(\ref{explicit-expression-Delta-R}) and (\ref{explicit-expression-Delta-Q}) to Eq.~(\ref{two-versions-NLPDA-equation}), we then obtain:
\begin{widetext}
\begin{small}
\begin{eqnarray}
- \frac{m}{4 \pi a_{F}} \, \Delta \! \left( \! \sqrt{\frac{x_{j}}{2 \lambda^{2}}} \right) & = & \frac{i^{l} \lambda^{D}}{\sqrt{\pi^{D}} \, y_{j}} \sum_{j'=1}^{N} \sum_{n'=0}^{N-1} (-1)^{n'} S^{T}_{j n'} S_{n' j'} \, y_{j'} 
\tilde{K} \! \left( \! \sqrt{\frac{x_{j'} \lambda^{2}}{2}} \Big| \sqrt{\frac{x_{j}}{2 \lambda^{2}}} \right) \tilde{\Delta} \! \left( \! \sqrt{\frac{x_{j'} \lambda^{2}}{2}} \right)
\label{algebraic-NLPDA-equation} \\
& = &  \frac{1}{y_{j}} \sum_{j'=1}^{N} \sum_{n'=0}^{N-1} (-1)^{n'} S^{T}_{j n'} S_{n' j'} \tilde{K} \! \left( \! \sqrt{\frac{x_{j'} \lambda^{2}}{2}} \Big| \sqrt{\frac{x_{j}}{2 \lambda^{2}}} \right) 
\sum_{j''=1}^{N} \sum_{n''=0}^{N-1} (-1)^{n''} S^{T}_{j' n''} S_{n'' j''} \, y_{j''} \Delta \! \left( \! \sqrt{\frac{x_{j''}}{2 \lambda^{2}}} \right) 
\nonumber 
\end{eqnarray}
\end{small}
\end{widetext}
\noindent
where $\sqrt{\frac{x_{j}}{2 \lambda^{2}}}$ stands for a value of $|\mathbf{r}|$ and $\sqrt{\frac{x_{j} \lambda^{2}}{2}}$ for a value of $|\mathbf{Q}|$ over the respective meshes of $N$ points.
Note that different symmetries enter Eq.~(\ref{algebraic-NLPDA-equation}) only through the index $l$ of the quantities $y_{j}$ (cf. Eq.~(\ref{definition-y})) and the index $\alpha = D/2+l-1$ of the $S$ matrix
(cf. Eq.~(\ref{S-matrix})).
Apart from this, the quantities $y_{j}$ and $S_{n j}$ in Eq.~(\ref{algebraic-NLPDA-equation}) are \emph{universal}, in the sense that they do not depend on coupling or temperature.
Note also that the first line of Eq.(\ref{algebraic-NLPDA-equation}) explicitly shows how the information on the gap parameter gets transferred from 
$\mathbf{Q}$- to $\mathbf{R}$-space and viceversa.
In the second line of Eq.(\ref{algebraic-NLPDA-equation}), on the other hand, this double transfer is embodied by the presence of four $S$ matrices.
In practice, we have found it convenient to solve the NLPDA equation using the version of the second line of Eq.(\ref{algebraic-NLPDA-equation}), where only $\Delta$ in real space appears explicitly.

For given values of coupling and temperature, Eq.~(\ref{algebraic-NLPDA-equation}) is then solved according to the following steps:
\vspace{-0.6cm}
\begin{enumerate}[(i)]
\item Choose a reasonable initial guess for $\Delta(r)$, to be inserted on the right-hand side of Eq.~(\ref{algebraic-NLPDA-equation}).
[For the isolated vortex with $D=2$ of subsection~\ref{sec:solving-NLPDA-equation}-A, we have taken as initial guess $\Delta(r) = \Delta_{0} r / \sqrt{1 + r^{2}}$
where $\Delta_{0}$ is the bulk value, which reproduces the expected behaviors at small and large $r$ (apart from numerical scaling factors).]
\vspace{-0.2cm}
\item On the basis of this guess, fix the initial values of $N$ and $\lambda$ by making a test on the (direct and inverse) Fourier transforms [cf. Eqs.~(\ref{explicit-expression-Delta-R}) and (\ref{explicit-expression-Delta-Q})].
\vspace{-0.2cm}
\item Calculate the values of $\Delta$ on the left-hand side of Eq.~(\ref{algebraic-NLPDA-equation}) over a coarse mesh of $x_{j}$ points (with $j=1,2,\dots,M$) where $M \ll N$ (typically, $M \approx 10^{2}$).
\vspace{-0.2cm}
\item At the next cycle of self-consistency, generate the values of $\Delta$, which are needed on the right-hand side of Eq.~(\ref{algebraic-NLPDA-equation}) over the fine mesh of $x_{j''}$ points (with $j''=1,2,\dots,N$), through a numerical interpolation on the values of $\Delta$ previously calculated on the coarse mesh of $M$ points.
\vspace{-0.1cm}
\item Repeat the process until self-consistency is attained.
[For the isolated vortex of subsection~\ref{sec:solving-NLPDA-equation}-A, typically $10$ cycles are sufficient.]
\vspace{-0.1cm}
\item Test the stability of the obtained self-consistent solution, by performing cycles of self-consistency with different values of $N$ and $\lambda$. 
[For the isolated vortex of subsection~\ref{sec:solving-NLPDA-equation}-A, values $N \approx 10^{3} \div 10^{4}$ and $\lambda \approx Q_{c}^{\mathrm{L}}$ of Eq.~(\ref{Q_c-Landau-criterion})
prove appropriate essentially for all couplings and temperatures.]
\end{enumerate}
\vspace{-0.2cm}

\begin{figure}[t]
\begin{center}
\includegraphics[width=8.5cm,angle=0]{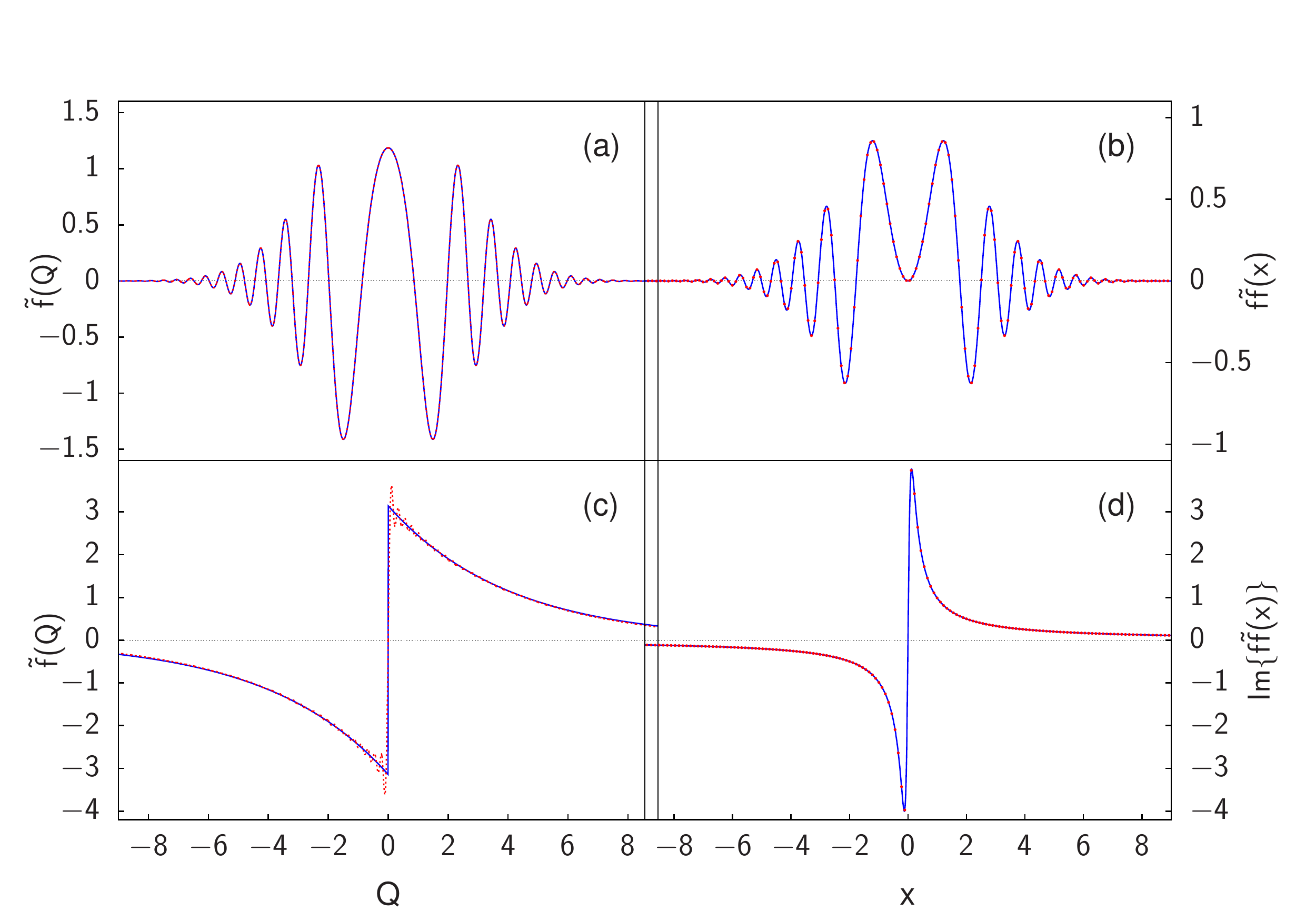}
\caption{(Color online) Comparison between the Fourier transform of the functions (\ref{test-function-1}) (a) and (\ref{test-function-2}) (c), as obtained analytically (full lines) 
                                     and by our numerical method (dashed lines). A similar comparison between the original functions (\ref{test-function-1}) and (\ref{test-function-2}) (full lines) 
                                     and their Fourier transforms taken twice (dots) is shown in (b) and (d).}
\label{Figure-19}
\end{center}
\end{figure}

There remains to show how the matrix elements $S_{n j}$ given by Eq.~(\ref{S-matrix}), which are needed in the expressions (\ref{explicit-expression-Delta-R}), (\ref{explicit-expression-Delta-Q}), and 
(\ref{algebraic-NLPDA-equation}), can be numerically generated in an efficient way.
The starting point is the following recursion relation valid for the (normalized) generalized Laguerre polynomials $\mathcal{L}^{\alpha}_{n}(u)$ that satisfy Eq.~(\ref{orthogonal-polynomials}) \cite{MOS-1966}:
\begin{eqnarray}
& & \sqrt{(n+1)(n+1+\alpha)} \, \mathcal{L}^{\alpha}_{n+1}(u) 
\label{recursion-relation} \\
& = & (2n+1+\alpha-u) \, \mathcal{L}^{\alpha}_{n}(u) - \sqrt{n(n+\alpha)} \, \mathcal{L}^{\alpha}_{n-1}(u) 
\nonumber
\end{eqnarray}
\noindent
where $n=1,2,\cdots$.
By cycling over this relation from $n=0$ up to $n=N$ and choosing for $u$ the $N$ values $\bar{u}$ such that $\mathcal{L}^{\alpha}_{N}(\bar{u})=0$ (corresponding to the $N$ distinct real zeros of the orthogonal 
polynomial $\mathcal{L}^{\alpha}_{N}(u)$), one ends up with the $N \times N$ eigenvalue problem:
\begin{widetext}
\begin{footnotesize}
\begin{equation}
\left(
\begin{array}{ccccc}
(1+\alpha) - \bar{u} \, , & - \sqrt{(1+\alpha)} \, , & 0 & \cdots & \cdots \\
- \sqrt{(1+\alpha)} \, , & (3+\alpha) - \bar{u} \, , & - \sqrt{2(2+\alpha)} \, , & 0 & \cdots \\
\cdots & \cdots & \cdots & \cdots & \cdots \\
\cdots & 0 & - \sqrt{(N-2)(N-2+\alpha)} \, , & (2N-3+\alpha) - \bar{u} \, , & - \sqrt{(N-1)(N-1+\alpha)} \\
\cdots & \cdots & 0 \, & - \sqrt{(N-1)(N-1+\alpha)} \, , & (2N-1+\alpha) - \bar{u} 
\end{array}
\right)
\left(
\begin{array}{c}
\mathcal{L}^{\alpha}_{0}(\bar{u}) \\
\mathcal{L}^{\alpha}_{1}(\bar{u}) \\
\cdots \\
\mathcal{L}^{\alpha}_{N-2}(\bar{u}) \\
\mathcal{L}^{\alpha}_{N-1}(\bar{u})
\end{array}
\right)
=
\left(
\begin{array}{c}
0\\
0 \\
0 \\
0 \\
0
\end{array}
\right) \, .
\label{big-eigenvalue-problem}
\end{equation}
\end{footnotesize}
\end{widetext}
\noindent
By diagonalizing the real and symmetric matrix that appears on the left-hand side of Eq.~(\ref{big-eigenvalue-problem}), one obtains the $N$ eigenvalues $\bar{u}_{j}$ (with $j=1,2,\cdots,N$) as well as the corresponding $N$ eigenvectors
\[
\begin{small}
\left(
\begin{array}{c}
\mathcal{L}^{\alpha}_{0}(\bar{u}_{j}) \\
\mathcal{L}^{\alpha}_{1}(\bar{u}_{j}) \\
\cdots \\
\mathcal{L}^{\alpha}_{N-2}(\bar{u}_{j}) \\
\mathcal{L}^{\alpha}_{N-1}(\bar{u}_{j})
\end{array}
\right) \, .
\end{small}
\]
\noindent
The ortho-normalization condition of these eigenvectors, namely,
\begin{equation}
\sum_{n=0}^{N-1} \mathcal{L}^{\alpha}_{n}(\bar{u}_{j}) \, \mathcal{L}^{\alpha}_{n}(\bar{u}_{j'}) = \frac{\delta_{j j'}}{w_{j}}
\label{normalization-eigenvectors}
\end{equation}
\noindent
then provides the factors $w_{j}$ according to the second of Eqs.~(\ref{orthogonality-S-matrix}).
The matrix elements of the $S$ matrix eventually result from their definition (\ref{S-matrix}).

We conclude this Appendix by presenting a few tests about the accuracy of the method we have developed to calculate numerically the Fourier transform of a function via 
Eqs.~(\ref{explicit-expression-Delta-R}) and (\ref{explicit-expression-Delta-Q}).
To this end, we consider two non-trivial (one even and one odd) functions, of the form
\begin{eqnarray}
f_{1}(x) & = & \exp{(-x^{2}/10)} \, \sin(x^{2})
\label{test-function-1} \\
f_{2}(x) & = & \frac{ i \, 64 x}{64 x^{2} + 1}
\label{test-function-2}
\end{eqnarray}
\noindent
whose Fourier transform can be obtained analytically by standard methods, yielding:
\begin{small}
\begin{eqnarray}
\tilde{f}_{1}(Q) & = & \sqrt{\frac{5 \pi}{101}} \, e^{-10 Q^{2}/101} \, \left[ \sqrt{ \sqrt{101}-1} \, \cos\left(\frac{100 Q^{2}}{101}\right) \right.
\nonumber \\
& - & \left. \sqrt{ \sqrt{101}+1} \, \sin\left(\frac{100 Q^{2}}{101}\right) \right]
\label{FT-test-function-1} \\
\tilde{f}_{2}(Q) & = & \pi \, e^{-|Q|/4} \, \sgn{(Q)} \, .
\label{FT-test-function-2}
\end{eqnarray}
\end{small}
\noindent
An additional test on the numerical method is obtained by calculating their Fourier transforms twice, thus returning back to the original functions (we have identified this operation by the
symbol $f\!\tilde{f}(x)$ to distinguish it from the original function $f(x)$).
Figures~\ref{Figure-19}(a) and (c) compare, respectively, the Fourier transforms of the test functions (\ref{test-function-1}) and (\ref{test-function-2}), as obtained analytically by the
expressions (\ref{FT-test-function-1}) and (\ref{FT-test-function-2}) (full lines) and by our numerical method (dashed lines) with $N=1000$ and $\lambda=1$.
In addition, Fig.s~\ref{Figure-19}(b) and (d) compare, respectively, the original test functions (\ref{test-function-1}) and (\ref{test-function-2}) (full lines) with the results of taking their Fourier transforms twice according to our numerical method (dots) (in Fig.~\ref{Figure-19}(d) the imaginary part of the test function (\ref{test-function-2}) 
has been reported).
In all cases, excellent agreement is obtained between the analytic and numerical calculations.
Note also the appearance of a Gibbs-like phenomenon, which occurs in Fig.~\ref{Figure-19}(c) at the edge of the discontinuity.



\begin{thebibliography}{99}
\bibitem{footnote-1} In Fig.~\ref{Figure-10}, the value of $R_{0}$ at $(k_{F}a_{F})^{-1}=0.5$ is somewhat larger than for other couplings. 
                                This is due to the procedure used to identify $R_{0}$, since close to $(k_{F}a_{F})^{-1}=0.5$ where the difference 
                                $K^{\sigma}(R_{0}) - K^{\sigma}_{\infty}(R_{0})$ changes sign (cf. Fig.~\ref{Figure-9}) the two kernels $K^{\sigma}(R)$ and $K^{\sigma}_{\infty}(R)$ 
                                run parallel to each other more than for other couplings. This results in a larger value of $R_{0}$ than for other couplings.
\bibitem{footnote-2} It is evident from the inset of Fig.~\ref{Figure-16} that, on the BCS side of unitarity, the values of $\xi$ obtained by the NLPDA and BdG calculations as well as those of
                                $\xi_{\mathrm{pair}}$ differ by constant factors.
                                These factors are (about) $\sqrt{2}$ when comparing the NLPDA and BdG calculations for $\xi$, and $\sqrt{3}$ when comparing $\xi_{\mathrm{pair}}$ with the BdG calculation for $\xi$.
\bibitem{footnote-3} In Fig.~\ref{Figure-18}, the values of $\delta_{Q}$ and $Q_{c}$ have been reported for $(k_{F}a_{F})^{-1} \le 0.5$ before reaching the coupling at which 
                                $\mu=0$ where $Q_{c}$ diverges.
                                Past this coupling, $\delta_{Q}$ is found to decrease as expected. For instance, $\delta_{Q}=(0.24,0.17)k_{F}$ for $(k_{F}a_{F})^{-1}=(1.0,1.5)$.
\end{thebibliography}
\end{document}